\documentstyle[12pt]{article}
\let\ssection=\section
\renewcommand{\section}{\setcounter{equation}{0}\ssection}

\newcommand\mathC{\mkern1mu\raise2.2pt\hbox{$\scriptscriptstyle|$}
        {\mkern-7mu\rm C}}          
\newcommand{\mathR}{{\rm I\! R}}         

\newtheorem{theorem}{Theorem}[section]

\newcommand\mapdown[1]{\Big\downarrow
        \rlap{$\vcenter{\hbox{$\scriptstyle#1$}}$}}
\newcommand\mapright[1]{\smash{
        \mathop{\mbox{\large{$\longrightarrow$}}}\limits^{#1}}}
\newcommand\bundle[3]{\begin{array}[t]{c}
        {#1}\\ \mapdown{#2}\\ {#3}\end{array}}
\newcommand\bundlemap[2]{\begin{array}[t]{c}
\mapright{#1}\\
\phantom{\mapdown{}}\\\mapright{#2}\\\end{array}}
\newcommand\I{(I)}

\begin{document}
\begin{titlepage}
\hspace{10truecm}Imperial/TP/97--98/71

\hspace{10truecm}quant-ph/9808067

\begin{center}
{\large\bf A Topos Perspective on the Kochen-Specker
                            Theorem:\\[6pt]
        II. Conceptual Aspects, and Classical Analogues}
\end{center}

\vspace{0.8 truecm}
\begin{center}
            J.~Butterfield\footnote{email:
            jb56@cus.cam.ac.uk;
            jeremy.butterfield@all-souls.oxford.ac.uk}\\[10pt]
            All Souls College\\
            Oxford OX1 4AL
\end{center}
\begin{center}
and
\end{center}

\begin{center}
            C.J.~Isham\footnote{email: c.isham@ic.ac.uk}\\[10pt]
            The Blackett Laboratory\\
            Imperial College of Science, Technology \& Medicine\\
            South Kensington\\
            London SW7 2BZ\\
\end{center}

\begin{center} 31 August 1998\footnote{Small corrections added
in October 1998} \end{center}

\begin{abstract}

In a previous paper, we have proposed assigning as the value
of a physical quantity in quantum theory, a certain kind of
set (a sieve) of quantities that are functions of the given
quantity. The motivation was in part physical---such a
valuation illuminates the Kochen-Specker theorem; and in part
mathematical---the valuation arises naturally in the topos
theory of presheaves.

This paper discusses the conceptual aspects of this proposal.
We also undertake two other tasks. First, we explain how the
proposed valuations could arise much more generally than just
in quantum physics; in particular, they arise as naturally in
classical physics. Second, we give another motivation for such
valuations (that applies equally to classical and quantum
physics). This arises from applying to propositions about the
values of physical quantities some general axioms governing
partial truth for any kind of proposition.

\end{abstract}
\end{titlepage}

\section{Introduction}\label{Sec:Intro}
In a previous paper \cite{IB98a}---referred to hereafter as
\I---we proposed assigning as the value of a physical quantity
in quantum theory, a certain kind of set (a sieve) of
quantities that are functions of the given quantity.  Our
motivation was in part physical---such a valuation illuminates
the Kochen-Specker theorem; and in part mathematical---the
valuation arises naturally in the topos theory of presheaves.
These aspects were closely linked. We interpreted a valuation
as assigning truth-values to propositions `$A\in\Delta$'
asserting that the value of the quantity $A$ lies in the Borel
subset $\Delta$ of the spectrum of the operator $\hat A$ that
represents $A$. The fact that one quantity can be a function,
or coarse-graining, of another implies that there is a natural
presheaf associated with these propositions. And the theory of
presheaves gives a natural generalization of the {\em FUNC\/}
property (viz. that the value of a function of a given
quantity is the function of the value of the quantity), which
plays a central role in the Kochen-Specker theorem.

In this paper, we first show how sieve-valued valuations
obeying our generalization of {\em FUNC\/} arise much more
generally than just as the values of quantities in quantum
physics; and accordingly, how the principal results of {\I}
can be generalized. In fact, we claim that they are one of the
most natural notions of valuation for any presheaf of
propositions, no matter what their topic. From a physical
perspective, a mathematical structure of this type is
indicated whenever the idea of `contextual' statements about
the system is (i) physically appropriate; and (ii) is so in
such a way that the set of all possible such contexts can be
regarded as the objects in a category, which then forms the
base category over which the presheaves are defined. As we
shall see, in making this claim we assume about valuations on
propositions only the basic idea that they must be some sort
of structure-preserving function from the set of propositions
(with operations such as negation, conjunction {\em etc.}
defined on it) to the set of truth values, which is to be some
sort of logical algebra.

That is the task of Section \ref{SieveValClassl}---where we
show that sieve-valued valuations arise naturally in classical
physics; and Section \ref{Sec:GenPropSieveValuations}---where
we show how such valuations can arise in even more general
contexts. But first, to facilitate reading the paper, there is
a short review of the elements of the theory of presheaves
(more concise than in \I, but with some extra heuristic
material), and of how these ideas were applied in {\I} to
quantum physics.

The paper ends with a presentation of another motivation for
such valuations (Section \ref{Sec:LogicPartialTruth}). We will
argue that intuitive ideas about what might be meant by the
notion of `partial truth' (applying to any type of
proposition) make sieve-valued valuations very natural.  Among
these principles, the main one will be that a proposition is
nearer to `total truth', the larger the subset of its
consequences that are themselves totally true. This argument,
and the principles it refers to, is conceptual, not
mathematical: indeed, it will not need the mathematical
notions of Section \ref{Sec:Prels}, except the idea of a
category---that {\em is} compulsory, in order to make sense of
the notion of a sieve. But the argument and its principles can
be made precise most naturally by using the ideas of presheaf
theory; in particular, the idea of `consequence' (entailment)
can be made precise in terms of the generalized notion of
coarse-graining introduced in Section
\ref{Sec:GenPropSieveValuations}.  Again, we shall see that
the proposals of {\I} arise from applying these general ideas
to propositions about the values of quantum physical
quantities.

We remark incidentally that there are still other motivations
for sieve-valued valuations obeying a generalized {\em
FUNC\/}. We discuss philosophical ones in \cite{BI98b}, and
physical ones in \cite{BI98c}, in each case adding further
material. For example, semantics for intuitionistic logic of
the Kripke-Beth type assigns to each formula as its
interpretation, a sieve on a poset; points of which are,
intuitively, possible states of knowledge, so that paths
represent possible courses of enquiry. In \cite{BI98b}, we
describe how this kind of construction suggests, as an
analogy, our own valuations.  In \cite{BI98c}, the motivation
concerns assigning to a quantity a Borel subset (rather than
an element) of its spectrum; it also is foreshadowed in \I.

\section{Review of Part I}
\label{Sec:Prels} In the first two subsections, we review
elements of the theory of presheaves. In the third, we
summarize how this theory was applied in {\I} to the
Kochen-Specker theorem, and to the idea of generalised
valuations on the physical quantities in a quantum theory.

\subsection{Categories, Presheaves and Subobjects}
\label{SubSec:cats&presheaves}

    A {\em presheaf\/} $\bf X$ on a small\footnote{A category is
said to be {\em small\/} if the collection of objects, and the
collection of all morphisms between a pair of objects, is a
set.} category $\cal C$ is a function that:
\begin{enumerate}
\item assigns to each $\cal C$-object $A$, a set
${\bf X}(A)$;

\item assigns to each $\cal C$-morphism $f:B\rightarrow A$,
a set-function, ${\bf X}(f):{\bf X}(A)\rightarrow {\bf X}(B)$;
and

\item makes these assignments in a `meshing' way, {\em i.e.},
${\bf X}({\rm id}_A)={\rm id}_{{\bf X}(A)}$; and, if
$g:C\rightarrow B$, and $f:B\rightarrow A$ then
\begin{equation}
    {\bf X}(f\circ g)={\bf X}(g)\circ{\bf X}(f)
                \label{Def:confunct}
\end{equation}
where $f\circ g:C\rightarrow A$ denotes the composition of $f$
and $g$.
\end{enumerate}
So intuitively, a presheaf is a collections of sets that vary
in a meshing way between `stages' or `contexts' $A,B,\ldots$
that are objects in the category $\cal C$. In terms of
contravariant and covariant functors, a presheaf on $\cal C$
is a contravariant functor from $\cal C$ to the category ${\rm
Set}$ of normal sets. Equivalently, it is a covariant functor
from the `opposite' category\footnote{The `opposite' of a
category $\cal C$ is a category, denoted ${\cal C}^{\rm op}$,
whose objects are the same as those of $\cal C$, and whose
morphisms are defined to be the same as those of $\cal C$ but
with each arrow reversed in direction.} ${\cal C}^{\rm op}$ to
${\rm Set}$.

To make the collection of presheaves on $\cal C$ into the
objects of a category, we recall that a morphism between two
presheaves $\bf X$ and $\bf Y$ is defined to be a {\em natural
transformation\/} $N:{\bf X}\rightarrow{\bf Y}$, by which is
meant a family of maps (called the {\em components\/} of $N$)
$N_A:{\bf X}(A)\rightarrow{\bf Y}(A)$, $A$ an object in $\cal
C$, such that if $f:B\rightarrow A$ is a morphism in $\cal C$,
then the composite map ${\bf X}(A)
\stackrel{N_A}\longrightarrow{\bf Y}(A)\stackrel{{\bf Y}(f)}
\longrightarrow{\bf Y}(B)$ is equal to ${\bf X}(A)
\stackrel{{\bf X}(f)}\longrightarrow{\bf X}(B)\stackrel{N_B}
\longrightarrow {\bf Y}(B)$. In other words, we have the
commutative diagram
\begin{equation}
    \bundle{{\bf X}(A)}{N_A}{{\bf Y}(A)}
    \bundlemap{{\bf X}(f)}{{\bf Y}(f)}
    \bundle{{\bf X}(B)}{N_B}{{\bf Y}(B)}    \label{cdNT}
\end{equation}
The category of presheaves on $\cal C$ equipped with these
morphisms is denoted ${\rm Set}^{{\cal C}^{\rm op}}$.

Since $\cal C$ is small, it follows that ${\rm Set}^{{\cal
C}^{\rm op}}$ is a topos. But we will not need the full
definition of a topos here\cite{Gol84,MM92}: it suffices that
it is a category that behaves much like the category ${\rm
Set}$, in particular as regards `subobjects'---the analogue of
subsets. To this we now turn.

\subsection{Subobjects, Sieves and Sections}
\label{SubSec:subobjects,sieves}
\paragraph*{1. Subobjects:}
The key idea about subobjects in a topos, which will be used
throughout this paper, is this. Just as an object in ${\rm
Set}$, {\em i.e.}, a set $X$, has subsets $K\subseteq X$ that
are in one-to-one correspondence with set-functions
$\chi^K:X\rightarrow\{0,1\}$, from $X$ to the special set
$\{0,1\}$, where $\chi^K(x)=1$ if $x\in K$, and $\chi^K(x)=0$
otherwise; so in a topos, the subobjects $K$ of an object $X$
are in one-to-one correspondence with morphisms
$\chi^K:X\rightarrow \Omega$, where the special object
$\Omega$ in the topos---called the `subobject
classifier'---forms an object of possible truth-values, just
as $\{0,1\}$ does in the category of sets.

We turn to the exact definitions. An object $\bf K$ is said to
be a {\em subobject\/} of $\bf X$ in the category of
presheaves if there is a morphism in the category ({\em i.e.},
a natural transformation) $i:{\bf K}\rightarrow {\bf X}$ with
the property that, for each stage $A$, the component map
$i_A:{\bf K}(A)\rightarrow{\bf X}(A)$ is a subset embedding,
{\em i.e.}, ${\bf K}(A)\subseteq {\bf X}(A)$.  Thus, if
$f:B\rightarrow A$ is any morphism in $\cal C$, we get:
\begin{equation}
    \bundle{{\bf K}(A)}{}{{\bf X}(A)}
    \bundlemap{{\bf K}(f)}{{\bf X}(f)}
    \bundle{{\bf K}(B)}{}{{\bf X}(B)}   \label{subobject}
\end{equation}
where the vertical arrows are subset inclusions. In
particular, it follows that ${\bf K}(f)$ is the restriction of
${\bf X}(f)$ to ${\bf K}(A)$.as

It is clear in what way the definitions above generalise the
ideas of set and subset. Namely, a presheaf over the category
$\cal C$ consisting of a {\em single\/} object $O$ corresponds
to a set $X:={\bf X}(O)$; and a subobject of this presheaf
corresponds to a subset of $X$.

\paragraph*{2. Sieves and the Subobject Classifier
${\bf \Omega}$:} To give the generalization for presheaves of
an ordinary subsets' characteristic function
$\chi^K:X\rightarrow\{0,1\}$, we first need the idea of a
sieve. A {\em sieve\/} on an object $A$ in $\cal C$ is defined
to be a collection $S$ of morphisms in $\cal C$ with codomain
$A$, and with the property that if $f:B\rightarrow A$ belongs
to $S$, and if $g:C\rightarrow B$ is any morphism, then
$f\circ g:C\rightarrow A$ also belongs to $S$.

With the idea of a sieve, one can immediately define the {\em
subobject classifier\/}. It is the presheaf ${\bf\Omega}:{\cal
C}\rightarrow {\rm Set}$ defined by:
\begin{enumerate}
    \item if $A$ is an object in $\cal C$, then
${\bf\Omega}(A)$ is the set of all sieves on $A$;

    \item if $f:B\rightarrow A$, then
${\bf\Omega}(f):{\bf\Omega}(A)\rightarrow{\bf\Omega}(B)$ is
defined as
\begin{equation}
    {\bf\Omega}(f)(S):= \{h:C\rightarrow B\mid f\circ h\in S\}
                                \label{Def:Om(f)}
\end{equation}
for all $S\in{\bf\Omega}(A)$; the sieve ${\bf\Omega}(f)(S)$ is
often written as $f^*(S)$, and is known as the {\em
pull-back\/} to $B$ of the sieve $S$ on $A$ by the morphism
$f:B\rightarrow A$.
\end{enumerate}

There are two main aspects to the idea that ${\bf\Omega}$
supplies an object of generalized truth-values. Both arise
from the basic idea mentioned in Section \ref{Sec:Intro}: that
a valuation on propositions (of any sort, not necessarily
about the values of physical quantities) must be some sort of
structure-preserving function from the set of propositions
(with some such operations as negation, conjunction etc.\
defined on it) to the set of truth values, which is to be some
sort of logical algebra.

The first aspect is the fact that for any $A$ in $\cal C$, the
set ${\bf\Omega}(A)$ of sieves on $A$ is a {\em Heyting
algebra}. This is a logical algebra that is distributive, but
more general than a Boolean algebra: the main difference being
in the behaviour of negation.  In this paper, we shall not
need the exact definition: we need only to remark that the
Heyting algebra structure of ${\bf\Omega}(A)$ is very natural;
and then to make an ensuing conceptual point.

Specifically, ${\bf\Omega}(A)$ is a Heyting algebra where the
partial ordering is defined on $S_1,S_2$ in ${\bf \Omega}(A)$
by $S_1\leq S_2$ if $S_1\subseteq S_2$; so that the unit
element $1_{{\bf\Omega}(A)}$ in ${\bf\Omega}(A)$ is the {\em
principal sieve\/} $\downarrow\!\!A := \{f:B\rightarrow A\}$
of all arrows whose codomain is $A$, and the null element
$0_{{\bf\Omega}(A)}$ is the empty sieve $\emptyset$. The
connectives for conjunction and disjunction are defined
as\footnote{The other key connective is the pseudo-complement
of $S_1$ relative to $S_2$. This is defined as $S_1\Rightarrow
S_2:=\{f:B\rightarrow A\mid \mbox{ for all $g:C\rightarrow B$
if $f\circ g\in S_1$ then $f\circ g\in S_2$}\}$. The negation
of an element $S$ is defined as $\neg S:=S\Rightarrow 0$; so
that $\neg S:=\{f:B\rightarrow A\mid \mbox{for all
$g:C\rightarrow B$, $f\circ g\not\in S$} \}$.}
\begin{eqnarray}
    && S_1\land S_2:=S_1\cap S_2    \label{Def:S1landS2}\\
    && S_1\lor S_2:=S_1\cup S_2.    \label{Def:S1lorS2}
\end{eqnarray}

The conceptual point is significant. It is that if for some
reason a set of propositions is associated with each $A$ in a
category $\cal C$ (perhaps, but not necessarily, as a
presheaf)---so that one is concerned to define {\em
contextual\/} valuations, {\em i.e.}, valuations associated
with each `context' or `stage of truth' $A$---then the set
${\bf\Omega}(A)$, being a Heyting algebra, forms a
`algebraically well-behaved' target space for such a valuation
associated with $A$.

The second aspect will be prominent in this Section and
beyond. It is the idea of generalizing to presheaves the way
that the subsets, $K$, of an ordinary set $X$ are in
one-to-one correspondence with characteristic functions
$\chi^K:X\rightarrow\{0,1\}$, from $X$ to the two classical
truth-values $\{0,1\}$. More precisely: the presheaf
${\bf\Omega}$ plays a role for the topos ${\rm Set}^{{\cal
C}^{\rm op}}$ analogous to the set $\{0,1\}$. That is to say,
subobjects of any object $\bf X$ in this topos ({\em i.e.},
subobjects of any presheaf on $\cal C$) are in one-to-one
correspondence with morphisms $\chi:{\bf X}\rightarrow
{\bf\Omega}$.

First, let $\bf K$ be a subobject of $\bf X$.  Then there is
an associated {\em characteristic morphism\/} $\chi^{{\bf
K}}:{\bf X}\rightarrow{\bf\Omega}$, whose component
$\chi^{{\bf K}}_A:{\bf X}(A)\rightarrow{\bf\Omega}(A)$ at each
$A$ in $\cal C$ is defined as
\begin{equation}
    \chi^{{\bf K}}_A(x):=\{f:B\rightarrow A\mid {\bf X}(f)(x)\in
                    {\bf K}(B)\} \label{Def:chiKA}
\end{equation}
for all $x\in {\bf X}(A)$.

That the right hand side of Eq.\ (\ref{Def:chiKA}) actually
{\em is\/} a sieve on $A$ follows from the defining properties
of a subobject. Thus, in each `branch' of the category $\cal
C$ going `down' from the stage $A$, $\chi^{{\bf K}}_A(x)$
picks out the first member $B$ in that branch for which ${\bf
X}(f)(x)$ lies in the subset ${\bf K}(B)$, and the commutative
diagram Eq.\ (\ref{subobject}) then guarantees that ${\bf
X}(f\circ h)(x)$ will lie in ${\bf K}(C)$ for all
$h:C\rightarrow B$. Hence each $A$ in $\cal C$ serves as a
possible `context' or `stage of truth' for an assignment to
each $x\in {\bf X}(A)$ of a generalised truth-value which is a
sieve, belonging to the Heyting algebra ${\bf\Omega}(A)$,
rather than an element of the Boolean algebra $\{0,1\}$ of
normal set theory.

Conversely, each morphism $\chi:{\bf X}\rightarrow{\bf\Omega}$
({\em i.e.}, each natural transformation between the
presheaves ${\bf X}$ and ${\bf\Omega}$) defines a subobject
${\bf K}^\chi$ of $\bf X$ by defining for each stage of truth
$A$
\begin{equation}
    {\bf K}^\chi(A):=\chi_A^{-1}\{1_{{\bf\Omega}(A)}\}=
    \{x\in{\bf X}(A)\mid \chi_A(x)=\downarrow\!\!A\}
                            \label{Def:KchiA}
\end{equation}
and by defining for each $f:B\rightarrow A$, the map ${\bf
K}^\chi(f): {\bf K}^\chi(A) \rightarrow {\bf K}^\chi(B)$ to be
the restriction of ${\bf X}(f)$ to ${\bf K}^\chi(A)$:
\begin{equation}
    {\bf K}^\chi(f):={\bf X}(f) |_{{\bf K}^\chi(A)}.
                            \label{Def:Kchif}
\end{equation}
Note that the fact that principal sieves pull back to
principal sieves ensures that Eq.\ (\ref{Def:Kchif}) implies
that, for any $x \in {\bf K}^\chi(A)$,
\begin{equation}
    \chi_{B}({\bf X}(f)(x)) = {\bf \Omega}(f)(\chi_{A}(x)) =
        {\bf \Omega}(f)(\downarrow\!\!A)=\downarrow\!\!B
\end{equation}
so that ${\bf X}(f)(x)\in{\bf K}^{\chi}(B)$, {\em i.e.}, ${\bf
K}^\chi$ is indeed a subobject of $\bf X$.

Note how this correspondence between subobjects and
characteristic morphisms simplifies in the special case
mentioned above (Section \ref{SubSec:subobjects,sieves}.1), of
presheaves on the category with a single object. In effect,
one gets just two truth-values---the unit element
$1_{{\bf\Omega}(O)}$ and the null element
$0_{{\bf\Omega}(O)}$, at the single stage of truth $O$; and
the component of the characteristic morphism at this single
stage is just the characteristic function of a subset of
$X:={\bf X}(O)$.

\paragraph*{3. Sections of a Presheaf:}
In any category, a {\em terminal object\/} is defined to be an
object $1$ such that, for any object $X$ in the category,
there is a unique morphism $X\rightarrow 1$. A {\em global
element\/} of an object $X$ is defined to be any morphism
$1\rightarrow X$. The motivation for this definition is that,
in the case of the category of sets, a terminal object is any
singleton set $\{*\}$; and then there is a one-to-one
correspondence between the elements of a set $X$ and functions
from $\{*\}$ to $X$.

For the category of presheaves on $\cal C$, a terminal object
${\bf 1}:{\cal C}\rightarrow {\rm Set}$ can be defined by
${\bf 1}(A):=\{*\}$ at all stages $A$ in $\cal C$; if
$f:B\rightarrow A$ is a morphism in $\cal C$ then ${\bf
1}(f):\{*\}\rightarrow\{*\}$ is defined to be the map
$*\mapsto *$. A global element of a presheaf $\bf X$ is also
called a {\em global section\/}. As a morphism $\gamma:{\bf
1}\rightarrow{\bf X}$ in the topos ${\rm Set}^{{\cal C}^{\rm
op}}$, a global section corresponds to a choice of an element
$\gamma_A\in{\bf X}(A)$ for each stage $A$ in $\cal C$, such
that, if $f:B\rightarrow A$, the `matching condition'
\begin{equation}
    {\bf X}(f)(\gamma_A)=\gamma_B \label{Def:global}
\end{equation}
is satisfied.

As discussed in the next Subsection, the Kochen-Specker
theorem is equivalent to the statement that certain presheaves
that arise naturally in quantum theory have no global
sections. But on the other hand, a presheaf may have
`partial', or `local', elements even if there are no global
ones. In general, a {\em partial element\/} of an object $X$
in a category with a terminal object is defined to be a
morphism $U\rightarrow X$, where $U$ is a subobject of the
terminal object $1$. In the category of sets, there are
no-nontrivial subobjects of $1:=\{*\}$, but this is not the
case in a general topos. In particular, in the case of
presheaves on $\cal C$, a partial element of a presheaf $\bf
X$ is an assignment $\gamma$ of an element $\gamma_A$ to a
certain {\em subset\/} of objects $A$ in $\cal C$---what we
shall call the {\em domain\/} ${\rm dom\,}\gamma $ of
$\gamma$---with the properties that (i) the domain is closed
downwards in the sense that if $A\in {\rm dom\,}\gamma$ and
$f:B\rightarrow A$, then $B\in {\rm dom\,}\gamma$; and (ii)
for objects in this domain, the matching condition Eq.\
(\ref{Def:global}) is satisfied.

\subsection{Some Applications to Quantum Physics}
\label{SubSec:applytoquantum} In this final Subsection, we
will illustrate how the notions reviewed in this Section can
be applied to the topic of valuations in quantum theory.
Again, we will be concise and pick out just some of the main
ideas of \I, leaving some to be generalized in later Sections,
and some wholly unmentioned.

\paragraph*{1. Categories of Quantities and the Kochen-Specker
Theorem:} We first introduce the set ${\cal O}_d$ of all
bounded self-adjoint operators with purely discrete spectra,
$\hat A,\hat B,\ldots$ on the Hilbert space $\cal H$ of a
quantum system.  We turn ${\cal O}_d$ into a category by
defining the objects to be the elements of ${\cal O}_d$, and
saying that there is a morphism from $\hat B$ to $\hat A$ if
there exists a real-valued function $f$ on $\sigma(\hat
A)\subset\mathR$, the spectrum of $\hat A$, such that $\hat
B=f(\hat A)$ (with the usual definition of a function of a
self-adjoint operator, using the spectral representation). If
$\hat B=f(\hat A)$, for some $f:\sigma(\hat
A)\rightarrow\mathR$, then the corresponding morphism in the
category ${\cal O}_d$ will be denoted $f_{{\cal O}_d}: \hat
B\rightarrow\hat A$.

We next form a presheaf on the category ${\cal O}_d$ from the
spectra of its objects. The {\em spectral presheaf} on ${{\cal
O}_d}$ is the covariant functor ${\bf\Sigma}:{{{\cal
O}_d}^{\rm op}}\rightarrow {\rm Set}$ defined as follows:
\begin{enumerate}
    \item On objects: ${\bf\Sigma}(\hat A):=\sigma(\hat
A)$---the spectrum of $\hat A$.

    \item On morphisms: If $f_{{\cal O}_d}:\hat B\rightarrow \hat A$,
so that $\hat B=f(\hat A)$, then ${\bf\Sigma}(f_{{\cal
O}_d}):\sigma(\hat A)\rightarrow \sigma (\hat B)$ is defined
by ${\bf\Sigma}(f_{{\cal O}_d})(\lambda):=f(\lambda)$ for all
$\lambda\in\sigma(\hat A)$.
\end{enumerate}

With these definitions, we can state one version of the
Kochen-Specker theorem in terms of presheaves. For recall that
one form of the theorem asserts that if $\dim{\cal H}>2$,
there does not exist an assignment $V$ to each object of
${\cal O}_d$ ({\em i.e.}, each bounded self-adjoint operator
on $\cal H$ with a discrete spectrum) of a member of its
spectrum, such that the so-called `functional composition
principle' (for short: {\em FUNC\/}) holds, viz. that for any
pair $\hat A$, $\hat B$ such that $\hat B= f(\hat A)$:
\begin{equation}
    V(\hat B)=f(V(\hat A))  .               \label{funct-rule}
\end{equation}
But this is precisely the `matching condition', Eq.\
(\ref{Def:global}), in the definition of a global element, as
applied to the spectral presheaf. Thus, in this form, the
Kochen-Specker theorem is equivalent to the statement that, if
$\dim{\cal H}>2$, there are no global elements of the spectral
presheaf ${\bf\Sigma}:{{\cal O}_d}^{\rm op}\rightarrow {\rm
Set}$. Note that we have restricted attention to operators
whose spectra are purely discrete on the grounds that it is
not physically meaningful to assign an {\em exact\/} value to
a quantity that lies in the continuous part of the spectrum of
the associated operator.

\paragraph*{2. From Partial Valuations to Generalised
Valuations:} Our next observation is that the Kochen-Specker
theorem permits the spectral presheaf to have partial
elements, as defined in Section
\ref{SubSec:subobjects,sieves}.3. In more usual, physical
language: it permits partial valuations, {\em i.e.}, an
assignment to each element $\hat A$ in some subset, ${\rm
dom\,}V$, of ${\cal O}_d$, of a member $V(\hat A)$ of $\sigma
(\hat A)$, such that: (i) ${\rm dom\,}V$ is closed under
taking functions of its members (`closed under
coarse-graining'); and (ii) for all $\hat A$, $\hat B$ $\in
{\rm dom\,}V$, with $\hat B= f(\hat A)$, {\em FUNC\/}, Eq.\
(\ref{funct-rule}), holds. And there are many such partial
valuations (whatever $\dim{\cal H}$). For example, each choice
of (i) an operator $\hat M$ with a purely discrete spectrum,
and (ii) one of its eigenvalues $m \in \sigma(\hat{M})$,
defines a partial valuation: one just takes ${\rm dom\,}V$ to
be the set of operators $\hat A$ that are functions of $\hat
M$, $\hat A=f(\hat M)$; and one defines $V(\hat M) := m$, and
$V(\hat A):=f(V(\hat M))=f(m)$.

The idea of a partial valuation brings us to our main claims
from {\I} (which remain central in this paper). There are in
effect three, which we will state briefly here, but explain in
order in this Paragraph and the next two.
\begin{enumerate}
\item Given such a partial valuation, there is a natural
associated valuation that:
    \begin{enumerate}
    \item is defined on {\em all\/} propositions `$A \in
\Delta$', stating that the value of the quantity $A$
(represented by the operator $\hat A$) lies in $\Delta$, a
Borel subset of $\sigma(\hat A)$;

    \item assigns to such a proposition as its value, a
sieve on $\hat A$ in the category ${\cal O}_d$.
\end{enumerate}
 These valuations have various properties, in particular an
analogue for sieves of the property {\em FUNC\/}: an analogue
that involves the idea of a pull-back.

\item {We then use these properties to generalise the notion
of a valuation. That is, we define a {\em generalised
valuation\/} as a map that (i) assigns a sieve on $\hat A$ to
each proposition `$A \in \Delta$', $\Delta \subseteq
\sigma(\hat A)$, and (ii) has these properties.

This definition has the desirable property that it can readily
be extended to the category ${\cal O}$ of {\em all\/} bounded
self-adjoint operators on the Hilbert space: specifically, a
proposition of the type `$A\in\Delta$' is physically (and
mathematically) meaningful irrespective of whether or not the
spectrum of $\hat A$ is purely discrete.
    }

\item We show that a quantum state (a pure state or a
mixture) defines such a generalised valuation in a natural
way.
\end{enumerate}

As to (1), the main idea is that---in the discrete case---even
if $\hat A$ is not in the domain, ${\rm dom\,}V$, of a partial
valuation $V$, for given $\Delta$ there might be one or more
functions $f$ such that (i) $f(\hat A)$ {\em does\/} lie in
${\rm dom\,}V$; and, (ii) $V(f(\hat A)) \in f(\Delta)$. This
situation prompts three observations.
\begin{itemize}
\item First: if a function $f$ satisfies conditions (i) and (ii),
then so does $g \circ f$ for any $g$; so the set of morphisms
in ${\cal O}_d$ associated with such functions determines a
{\em sieve\/} on $\hat A$ in ${\cal O}_d$.

\item Second: condition (ii) means that $V$ in effect assigns
{\em true\/} (in the usual classical sense!) to the
proposition `$f(A) \in f(\Delta)$'.

\item Third: the proposition `$f(A)\in f(\Delta)$' is weaker
than the original proposition `$A \in \Delta$', both
intuitively (since functions are generally not injective) and
mathematically, in the sense that its representing projector
is larger in the lattice of projectors on $\cal H$: ${\hat
E}[A \in \Delta] \leq {\hat E}[f(A) \in f(\Delta)]$.
\end{itemize}

Putting these observations together, we propose to assign to
`$A \in \Delta$' as a {\em contextual, partial\/} truth-value
at the stage $\hat A$, the sieve on $\hat A$ determined by the
functions obeying (i) and (ii). Formally, we define a {\em
generalised valuation\/} associated with a partial valuation
$V$ by
\begin{equation}
    \nu^V(A\in\Delta):=\{f_{{\cal O}_d}:
    \hat B\rightarrow \hat A
    \mid \hat B\in{\rm dom\,}V,\; V(\hat B)\in f(\Delta)\}.
                        \label{nVinDelta}
\end{equation}
This generalizes the values assigned by $V$ itself, in that
$V$'s assignments correspond to those propositions `$A \in
\Delta$' to which $\nu^V$ assigns the principal sieve,
$\downarrow\!\!\hat A := \{f_{{\cal O}_d}:\hat
B\rightarrow\hat A\}$, {\em i.e.}, the unit $1_{{\bf
\Omega}(\hat A)}$ of the Heyting algebra ${\bf \Omega}(\hat
A)$. We call this the {\em totally true\/} truth-value, ${\rm
true}_A$. Thus $\nu^V(A\in\Delta)={\rm true}_A$ if (i) $\hat
A$ lies in the domain of $V$;  and (ii) the value of $\hat A$
assigned by $V$ lies in the subset $\Delta\subseteq\sigma(\hat
A)$.

These definitions imply that the partial truth-value of `$A
\in \Delta$' at stage $\hat A$ is determined by those weaker
propositions `$f(A) \in f(\Delta)$' that are each totally true
at {\em their\/} stage $f(\hat A)$. For this partial
truth-value just {\em is\/} the sieve on $\hat A$ of
coarse-grainings $f(\hat A)$ of $\hat A$, at which `$f(A) \in
f(\Delta)$' is totally true. (We shall return to this idea in
Section \ref{Sec:LogicPartialTruth}.)

Generalised valuations associated with partial valuations have
various properties, of which we here mention just one, since
it will be significant in all later Sections (other properties
are listed in the next Paragraph). This property is the
analogue for sieves of the property {\em FUNC\/}. Roughly
speaking, it is that the value of a function of a quantity is
the pull-back of the quantity's value. To be precise: If
$f_{{\cal O}_d}:\hat B\rightarrow \hat A$, so that $\hat
B=f(\hat A)$, then
\begin{equation}
   \nu^V(B\in f(\Delta))=f_{{\cal O}_d}^*(\nu^V(A\in\Delta)).
                        \label{FUNCTpartval}
\end{equation}

    This property has two welcome consequences. First, we can
express the point of the previous paragraph in terms of
pull-backs. For note that for any category $\cal C$, with
objects $A,B,\ldots$, if $S$ is a sieve on $A$, and if
$f:B\rightarrow A$ belongs to $S$, then
\begin{equation}
    f^*(S):=\{h:C\rightarrow B\mid f\circ h\in S\}=
        \{h:C\rightarrow B\}= \ \downarrow\!\!B .   \label{f*S}
\end{equation}
Thus, for any category, the pull-back of a sieve on $A$ by any
morphism from $B$ to $A$ that belongs to the sieve, is the
principal sieve on $B$. Hence the pull-back of the truth-value
of `$A \in \Delta$' by a morphism within it, is total truth at
the context (stage of truth) that is the morphism's domain.
Second, there is a special, and especially intuitive, case of
the first point. Since, for any category, the pull-back of any
principal sieve by any morphism is the principal sieve, we can
say: if `$A \in \Delta$' is totally true (at stage $\hat A$),
then every weaker proposition `$f(A) \in f(\Delta)$' is
totally true (at its stage $f(\hat A)$).

\paragraph*{3. Generalised Valuations and the Coarse-Graining
Presheaf:} We turn now to claim (2), and use the various
properties possessed by generalised valuations associated with
partial valuations to define a wider notion of a {\em
generalised valuation\/} that is applicable to the category
$\cal O$ of all bounded, self-adjoint operators. Thus a
generalised valuation is defined to be any map that (i)
assigns a sieve on $\hat A$ to each proposition `$A \in
\Delta$'; and (ii) has these properties. For the sake of
completeness, we state these properties here; though in the
rest of this paper we shall only make substantial use of the
first---which is the sieve-analogue of {\em FUNC\/} (see {\I}
for the motivation for the other three):

\noindent {\em (i) Functional composition}:
\begin{eqnarray}
\lefteqn{\mbox{\ For any Borel function } f:\sigma(\hat A)
\rightarrow\mathR \mbox{ we have }}
\hspace{3cm}\nonumber\\[3pt]
    &&\nu(f(A)\in f(\Delta))=f_{\cal O}^*(\nu(A\in\Delta)).
\hspace{2cm} \ \label{FC-gen}
\end{eqnarray}

\noindent {\em (ii) Null proposition condition}:
\begin{equation}
    \nu(A\in\emptyset)=\emptyset=0_{{\bf \Omega}(A)} \label{Null-gen}
\end{equation}

\noindent {\em (iii) Monotonicity}:
\begin{equation}
\mbox{If }\Delta_1\subseteq\Delta_2\mbox{ then }
    \nu(A\in\Delta_1)\leq\nu(A\in\Delta_2)\mbox{,\ i.e.\ }
        \nu(A\in\Delta_1)\subseteq\nu(A\in\Delta_2).
                                    \label{Mono-gen}
\end{equation}

\noindent {\em (iv) Exclusivity}:
\begin{equation}
    \mbox{If $\Delta_1\cap\Delta_2=\emptyset$ and
$\nu(A\in\Delta_1)= {\rm true}_A$, then $\nu(A\in\Delta_2)<
{\rm true}_A$}. \label{Excl-gen}
\end{equation}

For later use in this paper, we also note that our collection
of sets of propositions `$A\in\Delta$' at each stage $\hat A$
can be made more precise; indeed, it can be regarded as a
presheaf, which we call the {\em coarse-graining presheaf\/}
$\bf G$ on $\cal O$. It is defined as follows:
\begin{enumerate}
    \item[(i)] For each $\hat A$ in $\cal O$, the set ${\bf
G}(\hat A)$ is defined to be the spectral algebra of $\hat A$,
{\em i.e.}, the algebra of spectral projectors ${\hat E}[A \in
\Delta]$ for the various Borel sets
$\Delta\subseteq\sigma(\hat A)$; thus, ${\bf G}(\hat A)$ can
be viewed as the Boolean algebra of all propositions of the
form `$A\in\Delta$'.

    \item[(ii)] For each morphism $f_{\cal O}:\hat B
\rightarrow\hat A$: the map ${\bf G}(f_{\cal O}): {\bf G}(\hat
A)\rightarrow{\bf G}(\hat B)$ is defined by
\begin{equation}
    {\bf G}(f_{\cal O})(\hat E[A\in\Delta]):= \hat E[f(A)\in
f(\Delta)]  \label{Def:G(fO)}
\end{equation}
or, equivalently, on propositions:
\begin{equation}
    {\bf G}(f_{\cal O})(\mbox{`$A\in\Delta$'}):=
            \mbox{`$f(A)\in f(\Delta)$'}.
\end{equation}
Note that the proposition `$f(A)\in f(\Delta)$' is equivalent
to the proposition `$A\in f^{-1}(f(\Delta))$', so the action
of ${\bf G}(f_{\cal O})$ can also be viewed as the explicit
coarse-graining operation
\begin{equation}
{\bf G}(f_{\cal O})(\mbox{`$A\in\Delta$'}):= \label{GfAD}
            \mbox{`$A\in f^{-1}(f(\Delta))$'}
\end{equation}
in which ${\bf G}(\hat B)$ is identified as the appropriate
subset of ${\bf G}(\hat A)$.
\end{enumerate}

In {\I} we remarked on the fact that, as it stands, the
right hand side of Eq.\ (\ref{Def:G(fO)}) is not
well-defined if the function $f$ and the Borel subset
$\Delta\subseteq\sigma(\hat A)$ are such that $f(\Delta)$
is not a {\em Borel\/} subset of $\mathR$. The way around
this is to note that if $f(\Delta)$ {\em is\/} a Borel
subset, then we have
\begin{eqnarray}
 \hat E[f(A)\in f(\Delta)]&=&\inf_{K\subseteq \sigma(f(\hat A))}
        \{\hat E[f(A)\in K]\mid \hat E[A\in\Delta]\leq
                    \hat E[f(A)\in K]\}\label{Def:f(A)inD}\\
    &=&\inf_{K\subseteq \sigma(f(\hat A))}
        \{\hat E[f(A)\in K]\mid \hat E[A\in\Delta]\leq
                    \hat E[A\in f^{-1}(K)]\}\\
    &=&\inf_{K\subseteq \sigma(f(\hat A))}
        \{\hat E[f(A)\in K]\mid \Delta\subseteq f^{-1}(J)\}
\end{eqnarray}
where the {\em infimum\/} is taken over all Borel subsets $J$
of $\sigma(f(\hat A))$. If $f(\Delta)$ is {\em not\/} a Borel
subset of $\mathR$, then we use Eq.\ (\ref{Def:f(A)inD}) as
the {\em definition\/} of $\hat E[f(A)\in f(\Delta)]$ for the
category of operators $\cal O$.

As we shall see in Section \ref{Sec:GenPropSieveValuations}
{\em et seq.\/}, the presheaf of propositions discussed above,
and its generalizations, play a central role in the
motivations for, and properties of, sieve-valued valuations
such as the generalised valuations just defined.

\paragraph*{4. Generalised Valuations from Quantum States:}
We turn to our claim (3) above: that quantum states naturally
define generalised valuations in the above sense. This
proceeds as follows.

The standard minimal interpretation of quantum theory holds
that a quantity $A$ possesses a value $a$ if, and only if, the
state $\psi$ is an eigenvector of $\hat A$ with eigenvalue
$a$; {\em i.e.}, $\hat A\psi=a\psi$: or more generally, that
`$A \in \Delta$' is true, if and only if, $\hat
E[A\in\Delta]\psi = \psi$. In terms of probability, it holds
that `$A \in \Delta$' is true if and only if ${\rm Prob}(A\in
\Delta;\psi)=1$ where ${\rm Prob}(A\in\Delta;\psi)$ denotes
the usual quantum mechanical (Born-rule) probability that the
result of a measurement of $A$ will lie in
$\Delta\subseteq\sigma(\hat A)\subset\mathR$, given that the
quantum state is $\psi$.

But in view of the discussion above, it is natural to reflect
that even if $\psi$ is not an eigenvector of $\hat A$, it is
an eigenvector of coarse-grainings $f(\hat A)$ of $\hat A$
(for example, the unit operator $\hat 1$ is always such a
function\footnote{If desired, and as explained in {\I}, such
`trivial' functions can be removed by replacing $\cal O$ with
the category ${\cal O}_*$: defined to be $\cal O$ minus all
real multiples of the unit operator.}); and hence we are lead
to propose that we should assign to the proposition `$A \in
\Delta$' the sieve of such coarse-grainings for which $\psi$
is in the range of the corresponding spectral projector ${\hat
E}[f(A) \in f(\Delta)]$.  Thus we define the generalised
valuation $\nu^\psi$ associated with a vector $\psi\in\cal H$
as
\begin{eqnarray}
    \nu^\psi(A\in\Delta)&:=&\{f_{\cal O}:\hat B\rightarrow\hat A
        \mid \hat E[B\in f(\Delta)]\psi=\psi\} \nonumber\\[2pt]
    &=&  \{f_{\cal O}:\hat B\rightarrow\hat A
        \mid {\rm Prob}(B\in f(\Delta);\psi)=1\}
                \label{Def:nupsiDelta}
\end{eqnarray}
where $\Delta$ is a Borel subset of the spectrum $\sigma(\hat
A)$ of $\hat A$. One can check that $\nu^\psi$ has all the
properties Eqs.\ (\ref{FC-gen}--\ref{Excl-gen}) required in
the definition of a generalised valuation.

Furthermore, one can give an exactly analogous definition of
the generalised valuation $\nu^\rho$ associated with a density
matrix $\rho$. One defines:
\begin{eqnarray}
    \nu^\rho(A\in\Delta)&:=&\{f_{\cal O}:\hat B\rightarrow \hat A
    \mid {\rm Prob}(B\in f(\Delta);\rho)=1\}\nonumber\\[2pt]
        &\,=&\{f_{\cal O}:\hat B\rightarrow \hat A
                \mid {\rm tr}(\rho\,\hat E[B\in f(\Delta)])=1\}.
                                         \label{Def:nurho}
\end{eqnarray}
Again, all the properties required of a generalised valuation
are satisfied.

\section{Sieve-valued Valuations in Classical Physics}
\label{SieveValClassl} Before developing the conceptual
aspects of the ideas of Section \ref{Sec:Prels} in a very
general setting (in the next Section), it is worth seeing how
they apply to classical physics. In the first Subsection, we
give a presheaf perspective on ordinary classical valuations.
We first make a category out of the real-valued functions on
phase space that represent classical physical quantities; we
then introduce the analogue of the spectral presheaf, and
contrast the classical existence of global sections with the
Kochen-Specker theorem; and we remark that quantisation can be
represented as a functor. In the second Subsection, we
motivate sieve-valuations in terms of classical macrostates.
In the third, we generalize this motivation, and present the
classical analogue of generalised valuations associated with a
partial valuation.

\subsection{A Presheaf Perspective on Orthodox Classical
Valuations} \label{PresheafPerspClassl} One usually thinks of
quantities and their values in classical physics as follows.
If $\cal S$ is the state space of some classical system, a
physical quantity $A$ is represented by a measurable
real-valued function $\bar A:{\cal S}\rightarrow\mathR$; and
then the value $V^s(A)$ of $A$ in any state $s\in\cal S$ is
simply
\begin{equation}
    V^s(A)=\bar{A}(s).                      \label{Vs(A)=A(s)}
\end{equation}
Thus all physical quantities possess a value in any state.
Furthermore, if $f:\mathR\rightarrow\mathR$ is a measurable
function, a new physical quantity $f(A)$ can be defined by
requiring the associated function $\overline{f(A)}$ to be
\begin{equation}
    \overline{f(A)}(s):=f(\bar{A}(s))       \label{Def:f(A)}
\end{equation}
for all $s\in\cal S$; {\em i.e.}, $\overline{f(A)}:=f\circ\bar
A:\cal S\rightarrow\mathR$. Thus by definition, the values of
the quantities $f(A)$ and $A$ satisfy a classical version of
{\em FUNC\/}:
\begin{equation}
    V^s(f(A))=f(V^s(A))                     \label{FUNCT-class}
\end{equation}
for all states $s\in\cal S$.

In terms of propositions of the form `$A\in\Delta$', where
$\Delta$ is a Borel subset of $\mathR$: to each microstate
$s\in\cal S$, there corresponds a valuation defined by
\begin{equation}
    V^s(A\in\Delta) =\left\{
        \begin{array}{ll}
            1 & \mbox{if $s\in \bar A^{-1}[\Delta]$} \\
            0 & \mbox{otherwise.}
        \end{array}
        \right.         \label{Def:Vs}
\end{equation}
Thus the proposition `$A\in\Delta$' is assigned the value
`true'(1) by $V^s$ if, and only if, $\bar A(s)\in\Delta$.

We turn now to rendering these ideas in terms of presheaves.
The notions introduced in the next two Paragraphs will also be
used in later subsections where we discuss sieve-valued
valuations for classical physics.

\paragraph*{1. The Category of Measurable Functions on $\cal S$:}
Let $\cal S$ be a classical state space, and let $\cal M$
denote the set of all real-valued measurable functions on
$\cal S$; thus each quantity $A$ is represented by
one\footnote{Strictly speaking, functions that differ only on
a set of Lebesgue measure zero should be identified; but we
shall ignore this subtlety here.} such function $\bar A:{\cal
S}\rightarrow\mathR$. We now regard $\cal M$ as a category
where: (i) the objects are the real-valued measurable
functions on $\cal S$; and (ii) we say there is a morphism
from $\bar B$ to $\bar A$ if there exists a measurable
function $f:{\cal S}(\bar A)\rightarrow\mathR$ such that $\bar
B=f\circ\bar A$ ({\em i.e.}, $\bar B(s)=f(\bar A(s))$, for all
$s\in\cal S$), where
\begin{equation}
    {\cal S}(\bar A):= \bar A({\cal S})= \{r\in\mathR\mid\exists
s\in{\cal S}, r=\bar A(s)\}
\end{equation}
is the set of all possible values that the physical quantity
$A$ could take; it is the classical analogue of the spectrum
$\sigma(\hat A)$ of a self-adjoint operator $\hat A$ in the
quantum theory . The morphism in $\cal M$ corresponding to
$f:{\cal S}(\bar A)\rightarrow\mathR$ will be denoted $f_{\cal
M}:\bar B\rightarrow \bar A$.

\paragraph*{2. The Value Presheaf:}
The analogue of the spectral presheaf in quantum theory is now
the following. We define the {\em value presheaf\/} on $\cal
M$ to be the covariant functor ${\bf \Upsilon}:{\cal M}^{\rm
op}\rightarrow {\rm Set}$ such that:
\begin{enumerate}
\item On objects:
${\bf\Upsilon}(\bar A):={\cal S}(\bar A)$---the set of all
possible values of the quantity $A$.

\item On morphisms: If $f_{\cal M}:\bar B\rightarrow\bar A$,
so that $\bar B=f\circ\bar A$, then ${\bf\Upsilon}(f_{\cal
M}):{\cal S}(\bar A)\rightarrow {\cal S}(\bar B)$ is defined
by ${\bf\Upsilon}(f_{\cal M})(\lambda):=f(\lambda)$ for all
$\lambda\in{\cal S}(\bar A)$.
\end{enumerate}

We now observe that a global section of the value presheaf
$\bf\Upsilon$ is a function $\gamma$ that assigns to each
object $\bar A$ in the category $\cal M$, an element
$\gamma_{\bar A}\in {\cal S}(\bar A)$ in such a way that if
$f_{\cal M}:\bar B\rightarrow\bar A$ (so that $\bar
B=f\circ\bar A$), then ${\bf\Upsilon}(f_{\cal M})(\gamma_{\bar
A})=\gamma_{\bar B}$; in other words
\begin{equation}
    \gamma_{\bar B}=f(\gamma_{\bar A}).
\end{equation}
Thus each global section corresponds to a classical valuation
that satisfies classical {\em FUNC\/}, as in Eq.\
(\ref{FUNCT-class}). Conversely, each such valuation
determines a global section of the value presheaf. Clearly,
the key difference from the situation in quantum theory (the
Kochen-Specker theorem) is that the classical presheaf {\em
does\/} have global sections: namely, each microstate
$s\in\cal S$, determines a global section $\gamma^s$ defined
by
\begin{equation}
        \gamma^s_{\bar A}:=\bar A(s)
\end{equation}
for all stages of truth $\bar A$.

\paragraph*{3. Quantisation as a Functor From $\cal M$ to
$\cal O$:} We remark incidentally that we can represent in
terms of $\cal M$ one of the main practical problems in
quantum physics; viz. knowing how to `quantise' a given
classical system. More precisely, one wants to associate to
each measurable function $\bar A:{\cal S}\rightarrow\mathR$, a
self-adjoint operator $\hat A$; or, perhaps, one seeks to do
this for some special subset of classical variables. There is
no universal way of performing such a quantisation; but it is
generally agreed that if a physical quantity represented by
$\bar A$ is associated in some way with a particular operator
$\hat A$, then, for any measurable function
$f:\mathR\rightarrow\mathR$, the function $f(\bar A)$ should
be associated with the operator $f(\hat A)$.

This preservation of functional relations can be represented
neatly in the language of category theory by saying that a
quantisation of the set of all classical quantities
corresponds to a {\em covariant functor\/} ${\bf Q}:{\cal
M}\rightarrow{\cal O}$ that is defined (i) on an object $\bar
A$ in $\cal M$ as ${\bf Q}(\bar A):=\hat A$; and (ii) on a
morphism $f_{\cal M}:\bar B\rightarrow\bar A$, by
\begin{equation}
        {\bf Q}(f_{\cal M}):=f_{\cal O}.
\end{equation}
This is because $f_{\cal M}:\bar B\rightarrow\bar A$, means
that $\bar B=f\circ\bar A$; and $f_{\cal O}:\hat
B\rightarrow\hat A$, means that $\hat B=f(\hat A)$.

\subsection{Motivating Sieve-valued Valuations for
Classical Physics} \label{SubSec:MotivateSieveValCP} Since
classical physics suffers no `Kochen-Specker prohibitions' on
global valuations of the orthodox kind, the motivation in
Section \ref{SubSec:applytoquantum}.2 for sieve-valued
valuations---as being naturally associated with {\em
partial\/} valuations---seems not to apply to classical
physics. But, in fact, the notion of a classical macrostate
motivates the classical analogue of a partial valuation, and
thereby leads to the associated sieve-valued valuations. This
Subsection describes the role of the notion of a macrostate;
and the next Subsection develops the idea so as to give the
exact classical analogue of the partial valuations discussed
in Section \ref{SubSec:applytoquantum}.2, and of the
associated generalised valuations.

So suppose we are given, not a microstate $s\in\cal S$, but
only a {\em macrostate\/}, represented by some Borel subset
$R\subseteq\cal S$: what then can be said about the `value' of
a quantity $A$, or the truth-value of a proposition `$A \in
\Delta$'? Various responses are possible\footnote{The most
familiar response is that in order to assign values, we must
do statistical physics; {\em i.e.}, we must have some
probability measure $\mu$ defined on $\cal S$, so that we say
the {\em probability\/} that the value of $A$ lies in
$\Delta$, given that the macrostate is $R$, is: $ {\rm
Prob}(A\in\Delta;R)= \mu(R\cap A^{-1}[\Delta])/\mu(R).$ But we
are asking what can be said about values, supposing we are
{\em not\/} doing statistical physics.}: for example, the
obvious choice is simply to say that the proposition
`$A\in\Delta$' is true in the macrostate $R$ if $\bar
A(R)\subseteq\Delta$, and false otherwise. Thus `$A\in\Delta$'
is defined to be true if, for {\em all\/} microstates $s$ in
$R$, the value $\bar A(s)$ lies in the subset $\Delta$.

However, one may feel that this assignment of true and false
is rather undiscriminating in so far as the proposition
`$A\in\Delta$' is adjudged false irrespective of whether $\bar
A(s)$ fails to be in $\Delta$ for all $s\in R$, or does so
only for a `few' points. For this reason, a more refined
response is to say that one wants the proposition
`$A\in\Delta$' to be `more true', the greater the set of such
points $s$: an idea that can be implemented by defining, for
example, a generalised truth-value $\upsilon^R(A\in\Delta)$ of
the proposition `$A\in\Delta$' to be the set of such points:
\begin{equation}
    \upsilon^R(A\in\Delta):=R\cap \bar A^{-1}[\Delta].
                            \label{Def:upR}
\end{equation}
Thus the set of possible truth-values of `$A\in\Delta$' is the
Boolean algebra of Borel subsets of $R\subseteq \cal S$; the
actual truth-value being the subset of $R$ in which the value
of $\bar A$ does belong to $\Delta$.  (So `totally true'
corresponds to the first response's `true', and is represented
by $R$ itself; whereas `totally false' is represented by the
empty set.) These two responses are certainly workable; we
discuss the second in another paper \cite{BI98b}.

But a third response is much more similar to what we have
discussed in the quantum case. Namely, we note that even if
$\bar A(R)$ is not a subset of $\Delta$, there will
be\footnote{The assertion `will be' is on the assumption that
constant functions on $\cal S$ are admitted. Such trivial
functions are the classical analogues of real multiples of the
unit operator $\hat 1$ in quantum physics, and---if
desired---they can be removed from the base category $\cal M$;
just as the quantum category $\cal O$ can be replaced with the
category ${\cal O}_*$ in which multiples of $\hat 1$ are
removed.} functions $f:\mathR\rightarrow\mathR$ with the
property that the `coarse-grained' function $f(\bar A):=f\circ
\bar A:{\cal S}\rightarrow\mathR$ satisfies the {\em weaker\/}
condition $f\circ \bar A(R)\subseteq f(\Delta)$; (so that,
according to the first response above, the weaker proposition
`$f(A)\in f(\Delta)$' {\em is\/} true in the macrostate $R$).
And we then define the generalised truth-value
$\nu^R(A\in\Delta)$ of the original proposition `$A\in\Delta$'
to be the set of all such coarse-grainings of $\bar A$.
Formally, in terms of the category $\cal M$:
\begin{equation}
        \nu^R(A\in\Delta):=\{f_{\cal M}:\bar B\rightarrow
            \bar A\mid \bar B(R)\subseteq f(\Delta)\}.
                \label{Def:nuR-rig}
\end{equation}
It is straightforward to check that the right hand side of
Eq.\ (\ref{Def:nuR-rig}) is a sieve on $\bar A$. Furthermore,
$\nu^R(A\in\Delta)$ has (classical analogues of) all the other
properties listed in Section \ref{SubSec:applytoquantum}.3 as
clauses of the general definition of a generalised
valuation---as we shall discuss in the next Subsection.

\subsection{The Classical Analogue of Generalised Valuations}
\label{SubSec:ClassGenVal} We will now generalize the use in
Section \ref{SubSec:MotivateSieveValCP} of macrostates to
motivate sieve-valued valuations, to obtain the classical
analogue of the generalised valuations in Section
\ref{SubSec:applytoquantum}.2 associated with {\em any\/}
partial valuation. All the properties discussed in Sections
\ref{SubSec:applytoquantum}.2 and
\ref{SubSec:applytoquantum}.3 (and incorporated as clauses of
the definition of a generalised valuation)---in particular,
the sieve-analogue of {\em FUNC\/}---will carry over to this
classical setting.

We begin by noting that a macrostate $R \subseteq\cal S$ is
naturally associated with a classical partial valuation $V^R$
({\em i.e.}, an assignment to some quantities of numbers as
values, obeying a classical version of {\em FUNC\/}). First we
define the domain of $V^R$ to be the set of all measurable
functions on $\cal S$ that are constant on the subset $R$:
\begin{equation}
    {\rm dom\,}V^R:=\{\bar A:{\cal S}\rightarrow \mathR\mid
            \forall s_1,s_2\in R, \bar A(s_1)=\bar A(s_2) \}.
\end{equation}
Then we define the value of a quantity $A$ whose
representative function $\bar A$ lies in the domain of ${\rm
dom\,}V^R$, by
\begin{equation}
            V^R(\bar A):=\bar A(s_0)
\end{equation}
for any $s_0\in R$; since $\bar A$ is constant on $R$, the
result does not depend on the choice of $s_0$ in $R$. So
$V^R(\bar A) \in {\cal S}(\bar A)$. It also follows that ${\rm
dom\,}V^R$ is closed under coarse-graining, and that the
values of $V^R$ obey {\em FUNC\/}. That is, we have, just as
in the definition of a partial valuation in Section
\ref{SubSec:applytoquantum}.2: if ${\bar A} \in {\rm
dom\,}V^R$ and $\bar B = f(\bar A)$ then (i) $\bar B \in {\rm
dom\,}V^R$; and (ii) $V^R(\bar B) = f(V^R(\bar A))$.

This prompts us to define a classical partial valuation in
general ({\em i.e.}, regardless of specifying a macrostate) as
an assignment $V$ to each element $\bar A$ belonging to some
subset ${\rm dom\,}V$ of $\cal M$, of a member of ${\cal S}
(\bar A)$, such that if $\bar B = f(\bar A)$ then (i) $\bar B
\in {\rm dom\,}V$ and (ii) $V(\bar B) = f(V(\bar A))$. With
this definition, claims (1) and (2) of Section
\ref{SubSec:applytoquantum}.2 and
\ref{SubSec:applytoquantum}.3 carry over completely to the
classical case: we simply substitute $\cal M$ for $\cal O$
(and so $\bar A$ for $\hat A$ etc.) and the `classical
spectrum' ${\cal S}(\bar A)$ for $\sigma(\hat A)$. Thus we
claim:
\begin{enumerate}
\item Given such a partial valuation $V$, there is a natural
 associated valuation that: (i) is defined on {\em all\/}
propositions `$A \in \Delta$'; and (ii) assigns to such a
proposition as its value, a sieve on $\bar A$ in the category
$\cal M$.  Namely:
\begin{equation}
\nu^{V}(A\in\Delta):=\{f_{\cal M}:\bar B\rightarrow \bar A\mid
\bar B\in{\rm dom\,}V, V(B)\in f(\Delta)\}.
                    \label{Def:nuR2Delta}
\end{equation}
Furthermore, the properties of these valuations, in particular
the analogue for sieves of the property {\em FUNC\/}, carry
over completely from the quantum to the classical case.

\item {Accordingly, we can use these properties to generalise
the notion of a valuation, {\em i.e.}, to define a {\em
generalised valuation\/} as a map that (i) assigns a sieve on
$\bar A$ to each proposition `$A \in \Delta$' and (ii) has
these properties.

    We can also present our collection of sets of propositions at
each stage $\bar A$ as a classical coarse-graining presheaf
$\bf G$ that (i) assigns to each $\bar A$ the Boolean algebra
of propositions of the form `$A\in\Delta$' (or, equivalently,
the algebra of characteristic functions $\chi_\Delta$: the
classical analogue of the spectral projectors in quantum
theory) identified as the Boolean algebra of Borel subsets
$\Delta \subseteq {\cal S}(\bar A)$; and (ii) acts on
morphisms $f_{\cal M}:B\rightarrow A$ such that ${\bf
G}(f_{\cal M})$ coarsens propositions, in exact analogy to
Eq.\ (\ref{GfAD}).  }
\end{enumerate}
But we shall not rehearse all the definitions, and
verifications of properties, substantiating these claims. For
firstly, they carry over directly from the discussion in {\I}
of the quantum case; and secondly, we shall see in the
Sections to follow that many of these definitions and
properties apply much more widely than in classical and
quantum physics.

Something that is worth developing a little further
however, is the observation that---as in the quantum
case---the situation can arise in which $f(\Delta)$ is not
a Borel subset of $\mathR$, even though $\Delta$ is. In
this context, we note that the central reason why it is
feasible to regard Eq.\ (\ref{Def:f(A)inD}) as a {\em
definition\/} of $\hat E[f(A)\in f(\Delta)]$ if $f(\Delta)$
is not Borel, is that the lattice of projection operators
is {\em complete\/}, and hence the right hand side of Eq.\
(\ref{Def:f(A)inD}) is well-defined. A natural analogue of
this construction in the classical case would be to start
with the Hilbert space $L^2(S,d\mu)$, where $d\mu$ is the
natural measure on the classical state space (a
$2n$-dimensional symplectic manifold, where $n$ is the
number of degrees of freedom) $S$ formed by taking the
wedge-product $n$-times of the basic symplectic $2$-form on
$S$. Any proposition $A\in\Delta$ can then be associated
with a corresponding {\em projection operator\/} on this
Hilbert space: namely, the projection onto the Borel subset
$A^{-1}(\Delta)$. An analogous trick to that in Eq.\
(\ref{Def:f(A)inD}) can then be applied by using the
projection lattice on the separable Hilbert space
$L^2(S,d\mu)$. However, we shall not go into the
mathematical details here since, in the present paper, our
invocation of the classical example is intended primarily
to be of pedagogical value as an illustration of the
general concepts that will be discussed in the next
Section.

 Finally, for the sake of
completeness, we remark on the classical analogue of claim
(3) of Section \ref{SubSec:applytoquantum}.4: the claim
that an orthodox quantum state, a vector $\psi \in \cal H$
or a density matrix $\rho$, induces a generalised
valuation. For $\psi\in\cal H$, we defined (see Eq.
(\ref{Def:nupsiDelta}))
\begin{eqnarray}
    \nu^\psi(A\in\Delta)&:=&\{f_{\cal O}:\hat B\rightarrow\hat A
        \mid \hat E[B\in f(\Delta)]\psi=\psi\} \nonumber\\[2pt]
    &=&  \{f_{\cal O}:\hat B\rightarrow\hat A
        \mid {\rm Prob}(B\in f(\Delta);\psi)=1\}\label{gladys}
\end{eqnarray}
where $\Delta$ is a Borel subset of the spectrum $\sigma(\hat
A)$ of $\hat A$. In the classical case, for $s \in \cal S$,
and $\Delta$ a Borel subset of the `classical spectrum' ${\cal
S}(\bar A)$, the analogue of Eq.\ (\ref{gladys}) is clearly
\begin{equation}
    \nu^s(A\in\Delta):=\{f_{\cal M}:\bar B\rightarrow\bar A
\mid \chi_{[B\in f(\Delta)]}(s)=1\} =
 \{f_{\cal M}:\bar B\rightarrow\bar A
    \mid f(\bar A(s))\in f(\Delta)\}\label{Def:nu.class.s.Delta}
\end{equation}
where $\chi_{[B\in f(\Delta)]}$ is the characteristic function
for ${\bar B}^{-1}(f(\Delta)$). It is easy to see that the
sieve $\nu^s(A\in\Delta)$ is the principal sieve, $1_{{\bf
\Omega}(A)} = \downarrow\!\!A$ (so that in the language of
Section \ref{SubSec:applytoquantum}.2, `$A \in \Delta$' is
totally true) if and only if $\bar A(s) \in \Delta$. One can
check that $\nu^s$ has all the properties required in the
definition of a generalised valuation (items (i) to (iv) in
Section \ref{SubSec:applytoquantum}.3).

Furthermore, there is a corresponding classical analogue of
the definition of the generalised valuation $\nu^\rho$
associated with a density matrix $\rho$:
\begin{eqnarray}
    \nu^\rho(A\in\Delta)&:=&\{f_{\cal O}:\hat B\rightarrow \hat A
\mid {\rm Prob}(B\in f(\Delta);\rho)=1\}    \\[2pt]
                   &\,=&\{f_{\cal O}:\hat B\rightarrow \hat A
\mid {\rm tr}(\rho\,\hat E[B\in f(\Delta)])=1\}.\nonumber
\end{eqnarray}
Namely: the classical analogue is that, with $\rho$ now
representing a classical mixed state, {\em i.e.}, a
probability measure on $\cal S$:
\begin{equation}
    \nu^\rho(A\in\Delta):=\{f_{\cal M}:\bar B\rightarrow \bar A
\mid {\rm Prob}_{\rho}(B\in f(\Delta))=1\}
                         \label{Def:nurhoClass}
\end{equation}
where ${\rm Prob}_{\rho}(B\in f(\Delta))$ is the classical
statistical probability, according to $\rho$, that `$B\in
f(\Delta)$', {\em i.e.}, the $\rho$-measure $\rho({\bar
B}^{-1}(f(\Delta)))$ of ${\bar B}^{-1}(f(\Delta))$. It follows
that the sieve $\nu^\rho(A\in\Delta)$ is the principal sieve,
$1_{{\bf \Omega}(A)} = \downarrow\!\! A$ if and only if
$\rho({\bar A}^{-1}(\Delta)) = 1$, {\em i.e. \/} if and only
if `$A \in\Delta$' is certain according to $\rho$.

\section{General Properties of Sieve-valued Valuations}
\label{Sec:GenPropSieveValuations} In this Section and the
next, we turn to showing how sieve-valued valuations arise
much more generally than just in the examples of quantum and
classical physics discussed earlier. Indeed, we claim that
they are one of the most natural notions of valuation for any
presheaf of propositions, no matter what their topic.  In
claiming this we will assume about valuations only the basic
idea that they must be some sort of structure-preserving
function from the sets of contextualised propositions (with
some such operations as negation, conjunction etc.\ defined on
it) to the corresponding sets of truth values, which are to be
some sort of logical algebra.

In this Section, we will argue for this claim by displaying
how some of the principal ideas and results of Section 4.2 and
Section 5 of {\I}---specifically, the sieve-version of {\em
FUNC\/} already emphasised in Sections \ref{Sec:Prels} and
\ref{SieveValClassl} above, and the notion of
`coarse-graining'---can be greatly generalized so that, for
the most part, they apply to {\em any\/} presheaf of
propositions. Another argument for the claim will be presented
in Section \ref{Sec:LogicPartialTruth} of the present paper.

\subsection{The Role of {\em FUNC\/}}
\label{SubSec:RoleFUNC} In this Subsection, we introduce our
most general version of {\em FUNC\/}; and motivate it and the
idea of a sieve-valued valuation on an arbitrary presheaf of
propositions $\bf G$, by showing that together they define
natural transformations from $\bf G$ to ${\bf\Omega}$, and
hence subobjects of $\bf G$.

Let $\cal C$ be any small category, with objects $A,B,\ldots$;
and let $\bf G$ be any presheaf on $\cal C$, with the set
${\bf G}(A)$ having elements $d,e,\ldots$. We think of the
pair $[A,d]$ as specifying a proposition at the context, or
stage of truth, $A$; and so of $\bf G$ as a presheaf of
propositions.  We call a function $\nu$ that assigns to each
choice of object $A$ and each $d \in {\bf G}(A)$, a set of
morphisms in $\cal C$ to $A$ ({\em i.e.}, morphisms with $A$
as codomain), a {\em morphism-valued valuation\/} on $\bf G$.
We write the values of this function as $\nu(A,d)$.

Note that for any set $S$ of morphisms to $A$ (not necessarily
a sieve), and any $f:B\rightarrow A$, we can define a
pull-back to $B$ of $S$ by Eq.\ (\ref{Def:Om(f)}):
\begin{equation}
    f^*(S):= \{h:C\rightarrow B\mid f\circ h\in S\}
                    \label{Def:f*S}
\end{equation}
although we note that there is no compelling reason for this
definition if the sets $S$ are totally unrestricted. However,
this caveat notwithstanding, we will say that a
morphism-valued valuation satisfies {\em generalized
functional composition\/}---for short, {\em FUNC\/}---if for
all $A,B$ and $f:B\rightarrow A$ and all $d\in{\bf G}(A)$, it
obeys
\begin{equation}
    \nu(B,{\bf G}(f)(d)) = f^*(\nu(A,d)).  \label{Def:GenlzdFunc}
\end{equation}

We call a morphism-valuation on $\bf G$ a {\em sieve-valued
valuation\/} on $\bf G$ if its values are all sieves; in this
case Eq.\ (\ref{Def:f*S}) is much better motivated since the
pull-back of a sieve is itself a sieve. The discussion in
Section \ref{Sec:Prels} already supplies us with two
motivations for using sieve-valuations in this very general
setting.  First, from a logical perspective: if we think of
${\bf G}(A)$ as a set of propositions, we expect a value
$\nu(A,d)$ of such a proposition to be some sort of
truth-value.  And we saw in Section \ref{Sec:Prels} how
$\bf\Omega$ supplies a well-behaved set of contextual and
generalized truth-values.

Second, and more generally: for any presheaf $\bf G$, a
natural notion of a valuation on $\bf G$ is a subobject of
$\bf G$. For think, as in logic, of a valuation as specifying
the `selected' or `winning' elements $d$ in each ${\bf G}(A)$.
One naturally imagines that these selected elements might form
a subobject of $\bf G$. But we saw in Section \ref{Sec:Prels}
that subobjects are in one-one correspondence with morphisms,
{\em i.e.}, natural transformations, $N:{\bf G}\rightarrow
{\bf\Omega}$. So one expects that at least some sieve-valued
valuations will define such a natural transformation by
$N^\nu_A(d):=\nu(A,d)$.

This motivation for sieve-valued valuations leads directly to
{\em FUNC\/}. For it turns out that {\em FUNC\/} is exactly
the condition a sieve-valued valuation must obey in order to
thus define a natural transformation, {\em i.e.}, a subobject
of $\bf G$. Specifically, we have ({\em cf.} Theorem 4.2 of
\I),

\begin{theorem}\label{genval=NT}
A sieve-valued valuation $\nu$ on $\bf G$ obeys {\em FUNC\/}
if and only if the functions at each stage of truth $A$
\begin{equation}
    N^\nu_A(d):=\nu(A,d)
\end{equation}
define a natural transformation $N^\nu$ from ${\bf G}$ to
${\bf\Omega}$.
\end{theorem}

\noindent{\bf Proof}\smallskip

\noindent Suppose $f:B\rightarrow A$, so that naturalness
means that the composite map ${\bf
G}(A)\stackrel{N^\nu_A}\longrightarrow{\bf\Omega}(A)
\stackrel{{\bf\Omega}(f)}\longrightarrow{\bf\Omega}(B)$ is
equal to ${\bf G}(A)\stackrel{{\bf G}(f)}\longrightarrow {\bf
G}(B)\stackrel{N^\nu_B}\longrightarrow {\bf\Omega}(B)$.  But
given that $N^\nu_A(d):=\nu(A,d)$, this is the condition that
\begin{equation}
    {\bf\Omega}(f)(\nu(A,d)) =
        (N^{\nu}_B\circ{\bf G}(f))(d) = \nu(B,{\bf G}(f)(d))
\end{equation}
which is exactly {\em FUNC\/}. \hfill {\bf Q.E.D.}

To sum up: we conclude that sieve-valued valuations obeying
{\em FUNC\/} are a very natural notion of valuation on any
presheaf of propositions

\subsection{Coarse-Graining Presheaves}
\label{SubSec:ValandCGPre} In this Subsection, we will
generalize one of the main notions in Section 5 of {\I}: the
idea of generalized coarse-graining.  Our generalization of
this notion involves the use of a new map, called the {\em
comparison functor\/}; this will also be needed in Section
\ref{Sec:LogicPartialTruth} in our general discussion of the
logic of partial truth.

There are two main ways in which we shall generalize the idea
of coarse-graining:
\begin{enumerate}

    \item In {\I}, the set of `propositions' ${\bf G}(\hat
A)$ at each stage $\hat A$ was a Boolean algebra (of Borel
subsets of $\sigma(\hat A)$, or equivalently of $\hat A$'s
spectral projectors; and similarly for the classical case,
{\em cf.\/} Section 3). However, here we shall assume only
that ${\bf G}(A)$ is a poset with a $0$ and a $1$. Indeed,
much of what follows could be generalized to the case where
${\bf G}(A)$ is just a poset; but we will also require a $0$
and a $1$, to link to the null, exclusivity and monotonicity
clauses of the definition of a generalised valuation, Eqs.\
(\ref{Null-gen}--\ref{Excl-gen}) (see Eqs.\
(\ref{Def:locval1}--\ref{Def:locval3}) below).

\item  In \I,  generalized coarse-graining was defined so as
to use, for the case where $f_{\cal O}:{\hat B}\rightarrow
{\hat A}$, the identity map on ${\bf G}(\hat B)$ to embed the
Boolean algebra ${\bf G}(\hat B)$ into its superset (larger
Boolean algebra) ${\bf G}(\hat A)$ (this was used in writing
Eq.\ (\ref{GfAD})).  In the generalization in this Subsection
to any presheaf of propositions ${\bf G}$ on any small
category $\cal C$, such that ${\bf G}(A)$ is a poset with a
$0$ and $1$, we will again need a map acting in the opposite
direction to ${\bf G}(f)$. But it need not be the identity
map, since the poset ${\bf G}(B)$ need not be a subset of
${\bf G}(A)$.  So we will simply assume that there is some
such map (given by the {\em comparison\/} functor introduced
in Paragraph 1 below).
\end{enumerate}

We should also note another way in which the exposition to
follow differs from that in Section 5 of \I. There, our
discussion took as the base-category, not $\cal O$, but the
poset $\cal W$ of all Boolean subalgebras of the projection
lattice ${\cal P}({\cal H})$ of the Hilbert space $\cal H$. In
this category\footnote{In a similar way, one can make a
category out of any poset; in particular, the corresponding
category for classical physics will consist of all Boolean
subalgebras of the algebra (itself Boolean!) of all Borel
subsets of the classical state space $\cal S$, again ordered
by subalgebra inclusion.}, the objects are defined to be the
subalgebras $W\in\cal W$; and a morphism is defined to exist
from $W_2$ to $W_1$ if $W_2\subseteq W_1$: thus there is at
most one morphism between any two objects. In some respects
$\cal W$ is a more natural category to work with than $\cal
O$, since it `identifies' quantities that are each a function
of the other, and hence have the same spectral algebra.

For this reason, in {\I} we sometimes worked with $\cal W$,
rather than $\cal O$. In particular, the Kochen-Specker
theorem gets as natural an expression in terms of $\cal W$, as
it does in terms of $\cal O$. But, for the sake of brevity, in
the review of {\I} in the present paper we have used only the
category $\cal O$ (and its classical analogue $\cal M$). And
again in this Subsection, while generalizing Section 5 of
{\I}, we will present our definitions and results in terms of
the category $\cal C$, which we have hitherto thought of as
generalizing $\cal O$. So for the rest of this Subsection, we
assume as in Section \ref{SubSec:RoleFUNC} that $\cal C$ is
any small category, with objects $A,B,\ldots$; and that $\bf
G$ is a presheaf on $\cal C$, with the set ${\bf G}(A)$ having
elements $d,e,\ldots$. We also assume that at each $A$, ${\bf
G}(A)$ is a poset with a $0$ and a $1$.

\paragraph*{1. The Comparison Functor:}

In Section \ref{Sec:LogicPartialTruth} below, given a morphism
$f:B\rightarrow A$, we shall need to be able to `push-forward'
a proposition $d \in {\bf G}(B)$ to ${\bf G}(A)$, for
comparison of `logical strength' ({\em i.e.}, comparison
according to the partial order $<$ in the poset ${\bf G}(A)$)
with propositions in ${\bf G}(A)$.

In {\I}, this presented no problem since, in using the base
category $\cal W$, we have that $d \in {\bf G}(B)(=W_2)$ is
itself also a member of ${\bf G}(A)(=W_1)$, (if
$f:W_2\rightarrow W_1$, so that $W_2\subseteq W_1$). But with
a general category $\cal C$, this fails since there is no {\em
a priori\/} embedding of ${\bf G}(B)$ in ${\bf G}(A)$.

Accordingly, we now assume that such a map is given. More
precisely, we assume that whenever $f:B\rightarrow A$ in $\cal
C$, we are given a map from ${\bf G}(B)$ to ${\bf G}(A)$,
which need not be injective. For much of the argument to
follow, we do {\em not\/} need to assume that these maps mesh
under composition so as to give a (covariant) functor from
$\cal C$ to $\rm Set$, but for simplicity, we will do so. Thus
we assume that there is a covariant functor, $\bf C$, from
$\cal C$ to $\rm Set$, called the {\em comparison functor\/}
(`C' for `comparison'), with the same action on objects $A$ in
$\cal C$ as has $\bf G$. To sum up:
\begin{itemize}
\item ${\bf C}(A):={\bf G}(A)$ at all $A$;

\item if $f:B\rightarrow A$, there is a map
${\bf C}(f): {\bf C}(B)\rightarrow {\bf C}(A)$.
\end{itemize}

\paragraph*{2. Coarse-Graining Presheaves:}
We turn now to the main topic of this Subsection, which is to
generalize the discussion in \I, Section 5, of generalized
coarse-graining. In effect, that discussion proceeded by
noting three properties of the original coarse-graining
presheaf ${\bf G}:{\cal O}\rightarrow{\rm Set}$ (defined in
Section \ref{SubSec:applytoquantum}.3 above); and then
defining {\em a\/} coarse-graining presheaf to be any presheaf
with these properties. These properties were called
`coarse-graining', `retraction' and `monotonicity'; but we
need not list them. (We say `in effect', just because the
definition was in terms of the category $\cal W$, not $\cal
O$.)

Here, we will generalize to any small category $\cal C$. The
idea is to take the comparison functor $\bf C$ to be given
{\em ab initio\/}, and then to define a presheaf $\bf G$ to be
a `coarse-graining' with respect to $\bf C$, if it has these
three properties---or rather, their generalizations, to allow
for ${\bf C}(f)$ not necessarily being a subset inclusion map.

So we assume we are given a covariant functor $\bf C$ from
$\cal C$ to $\rm Set$, with all the ${\bf C}(A)$ being posets
with a $0$ and $1$. Then we define a {\em coarse-graining with
respect to\/} $\bf C$ to be a presheaf $\bf G$ on $\cal C$
({\em i.e.}, a contravariant functor from $\cal C$ to $\rm
Set$), with the following properties:
\begin{enumerate}
    \item[1)] $\bf G$ has the same action on objects as $\bf
    C$, {\em i.e.}, ${\bf G}(A) := {\bf C}(A)$;

    \item[2)] `coarse-graining': if $f:B\rightarrow A$, then for
all $d \in {\bf G}(A)$,
\begin{equation}
        d \leq {\bf C}(f)[{\bf G}(f)(d)];\label{Def:CoarseGrain}
\end{equation}

    \item[3)] `monotonicity': if $f:B\rightarrow A$, and $d \leq
e$ in ${\bf G}(A)$, then ${\bf G}(f)(d) \leq {\bf G}(f)(e)$ in
${\bf G}(B)$.
\end{enumerate}
In {\I} we also added the condition
\begin{enumerate}
\item[4)] `generalized retraction': if $f:B\rightarrow A$,
then for all $d \in {\bf G}(B)$,
\begin{equation}
    {\bf G}(f)[{\bf C}(f)(d)] = d       \label{Def:GenRetr}
\end{equation}
\end{enumerate}
but we note that if this extra condition Eq.\
(\ref{Def:GenRetr}) is imposed, then the map ${\bf C}(f):{\bf
G}(B)\rightarrow {\bf G}(A)$ is necessarily injective ({\em
i.e.}, it is one-to-one); and hence we have only a marginal
generalisation of the situation in {\I} in which ${\bf G}(B)$
is an explicit subset of ${\bf G}(A)$.  On the other hand, the
motivation for imposing the generalised retraction condition
in the first place was closely linked to the fact that ${\bf
G}(B)$ {\em is\/} a subset of ${\bf G}(A)$ in the example of
quantum theory; therefore it is legitimate to consider
removing this condition, with a concomitant freeing up of
possibilities for the comparison-functor maps ${\bf C}(f):{\bf
G}(B)\rightarrow{\bf G}(A)$.

\paragraph*{3. Generalised Valuations on a general
Coarse-graining Presheaf:} The general notion of a
coarse-graining presheaf just introduced admits generalised
valuations of the {\em FUNC\/}-obeying kind originally
envisaged in {\I} and in Section \ref{SubSec:RoleFUNC}.

The first step is to define a {\em local valuation\/} of the
poset ${\bf G}(A)$ in the Heyting algebra ${\bf \Omega}(A)$.
This is to be a map $\phi:{\bf G}(A)\rightarrow
{\bf\Omega}(A)$ such that the following conditions are
satisfied:
\begin{eqnarray}
    &&{\rm Null\ proposition\ condition:}\quad
        \phi(0_{{\bf G}(A)})=0_{{\bf\Omega}(A)}
                                \label{Def:locval1}\\[3pt]
    &&{\rm Monotonicity:}\quad \alpha\leq\beta\mbox{ implies }
                    \phi(\alpha)\leq\phi(\beta)
                                \label{Def:locval2}\\[3pt]
    &&{\rm Exclusivity:}\quad
        \mbox{ If }\alpha\land\beta=0_{{\bf G}(A)}
            \mbox{ and }\phi(\alpha)=1_{{\bf\Omega}(A)},
                \mbox{ then }\phi(\beta)<1_{{\bf\Omega}(A)}
                                \label{Def:locval3}
\end{eqnarray}
which are the appropriate analogues of Eq.\ (\ref{Null-gen}),
Eq.\ (\ref{Mono-gen}) and Eq.\ (\ref{Excl-gen}) respectively.

    Now we define a {\em generalised valuation\/} on $\cal
C$ associated with a coarse-graining presheaf $\bf G$ ($\bf G$
being with respect to some comparison functor $\bf C$) to be a
family of local valuations $\phi_A:{\bf G}(A)\rightarrow
{\bf\Omega}(A)$, at each $A$, such that if $f:B\rightarrow A$
then, for all $d \in {\bf G}(A)$,
\begin{equation}
    \phi_{B}({\bf G}(f)(d)) = f^*(\phi_{A}(d)).
\end{equation}
Bearing in mind that this equation is essentially {\em FUNC},
Eq.\ (\ref{Def:GenlzdFunc}), and that local valuations obey
the null proposition, monotonicity and exclusivity conditions
in Eqs. (\ref{Def:locval1}--\ref{Def:locval2}), we see that
this definition directly generalizes the generalised
valuations on $\cal O$ of Section
\ref{SubSec:applytoquantum}.3. So the definition is non-empty.
In particular: in \I, Section 5.3.4, we showed that a density
matrix defines such a generalised valuation on the specific
category $\cal W$, associated with any coarse-graining
presheaf on $\cal W$. A similar result can be proved for the
classical case, using the material at the end of Section 3
above, especially Eq.\ (\ref{Def:nurhoClass}).

Finally, we remark that since these generalised valuations for
an arbitrary coarse-graining presheaf $\bf G$ (with respect to
an arbitrary comparison functor) obey {\em FUNC\/}, the
discussion of Section \ref{SubSec:RoleFUNC} applies. That is:
each such generalised valuation, $\Phi$ say, (a family of
local valuations $\phi_A$) defines a natural transformation
$N^{\Phi}$ from the coarse-graining presheaf $\bf G$ to the
subobject classifier $\bf \Omega$, by defining the components:
\begin{equation}
    N^{\Phi}_{A}(d) := \phi_{A}(d)
\end{equation}
As emphasised in Section \ref{SubSec:subobjects,sieves}, such
natural transformations are in one-to-one correspondence with
subobjects. Thus each such generalised valuation defines a
subobject of $\bf G$.

\section{The Logic of Partial Truth}
\label{Sec:LogicPartialTruth} We turn now to give our final
motivation for the use of sieve-valued valuations. We start
from a handful of general intuitive requirements about how the
truth-values of propositions should reflect their logical
relations, and argue that sieve-valued valuations are the
natural way to satisfy these requirements. More precisely:
valuations taking sieves as their values are determined in a
natural way, for any category $\cal C$ of `contexts', once we
require the following:
\begin{enumerate}
    \item[(i)] each object in the category has an associated
family of propositions, with different families corresponding
to different objects families meshing suitably;

    \item[(ii)] the valuation is to represent partial truth
(degrees of truth), subject to some weak conditions, the most
important being that the partial truth-value of a proposition
at a stage $A$ in $\cal C$ is to be determined by which of its
consequences (weaker propositions) are totally true at their
own stage.
\end{enumerate}
The concrete valuations discussed in Sections
\ref{SubSec:applytoquantum} and \ref{SieveValClassl} (and in
\I) arise from applying these requirements to propositions
about the values of physical quantities.

We emphasise that although we think conditions in (ii) on
partial truth are very reasonable, we make no claim that they
are obligatory. In the philosophical literature, partial truth
is modelled in various ways, and indeed often rejected
altogether (for example, \cite{Haa80}). We discuss this more
in \cite{BI98b}. Here, suffice it to say in defence of our own
notion that at least it is tightly controlled by the notion of
total truth, in the sense that the partial truth-value of any
proposition is determined by which propositions are totally
true.

Our argument will be very general: indeed, the only precise
mathematical notion that is needed is that of a sieve in a
category. Otherwise, the argument can be formulated
intuitively: for example, in its use of the idea of one
proposition being a consequence of (logically weaker than)
another. Of course, by assuming mathematical notions in
addition to that of a category, these intuitive ideas can be
made precise. But it seems to us best to emphasise the
generality of the intuitive argument by assuming these further
notions only after giving the argument.

We will therefore proceed in two subsections. Subsection 1
will give the intuitive argument that assumes only the notion
of a category, and leads to sieve-valued valuations.
Subsection 2 will comment on the argument, and exhibit one
natural way of making its intuitive ideas precise: in
particular, making consequence (entailment) precise by having
the families of propositions at each stage be posets, and
having embedding maps like the comparison functor introduced
in Section \ref{SubSec:ValandCGPre}.

\subsection{The Intuitive Argument for Sieve-valued Valuations}
\label{SubSec:IntuitiveArgument} Suppose we are given some
category $\cal C$, and that to each object $A \in \cal{C}$ is
associated a set ${\cal P}(A)$ whose elements $d$ we will call
`items'. We allow that for different objects \mbox{$A,B$ in
$\cal{C}$}, the sets ${\cal P}(A)$, ${\cal P}(B)$ can differ.
For each $A$ and $d \in$ ${\cal P}(A)$ we think of $[A,d]$ as
a proposition. We do not require that for fixed $A$, the
family $\{[A,d]\mid d\in{\cal P}(A)\}$ is a Boolean algebra;
nor, for the moment, that it have any other structure---for
example, that of a poset. But we do require the following
assumptions.

\begin{enumerate}
    \item[(A)] {The morphisms in the category are associated with
maps between propositions for different objects, as follows.
If there is a morphism $f:B \rightarrow A$ from $B$ to $A$,
then there is a function from the family of propositions
$\{[A,d]\mid d\in{\cal P}(A)\}$ to the corresponding family
$\{[B,e]\mid e\in{\cal P}(B)\}$ associated with $B$. We
represent this map associated with $f$ by $f^\dagger$ acting
on the items $d$. So given a morphism $f:B \rightarrow A$,
then $[B,f^\dagger(d)]$ is the `$B$-proposition' that
`corresponds by $f$' to $[A,d]$. Furthermore, recalling that
every object $A$ in a category has an identity morphism, ${\rm
id}_{A}:A \rightarrow A$, we require that the map on
propositions associated with the identity morphism be the
identity map on propositions. That is: we require that for any
$A$, $({\rm id}_{A})^\dagger = {\rm id}_{{\cal P}(A)}$. Two
remarks about this assumption.
\begin{enumerate}

    \item {We do not initially require that the
associations $A\mapsto {\cal P}(A)$ and $f \mapsto f^\dagger$
together define a presheaf on $\cal C$.  That is: we do not
need to assume that, given morphisms $f:B \rightarrow A$ and
$g:C \rightarrow B$, and so a morphism $f \circ g: C
\rightarrow A$, we have: $g^\dagger (f^\dagger (d)) = (f \circ
g)^\dagger(d)$.

    However, we note {\em en passant\/} that if this presheaf
condition is {\em not\/} satisfied, then the
`$\dagger$'-operation is `path-dependent' in the following
sense: If a morphism $k:C\rightarrow A$ can be factored in the
form $C\stackrel g\rightarrow B\stackrel f\rightarrow A$, then
the pull-back $k^\dagger(d)$ of $d\in{\cal P}(A)$ may not
equal the composite pull-back $g^\dagger(f^\dagger(d))$
obtained by factoring $k$ through the intermediate object $B$.
In most physical situations, such a behaviour would be
considered distinctly pathological.  }

    \item To use the notation ${\bf G}(f)$ instead of $f^\dagger$
would echo the notation in Section \ref{SubSec:ValandCGPre}
(and its special cases, Definitions 5.3 and 5.4  of \I).  But
we use $f^\dagger$ to indicate that we do not require a
presheaf. See the next subsection for how the argument to come
can be carried over to any coarse-graining presheaf in the
sense of Section \ref{SubSec:ValandCGPre}.
\end{enumerate}
    }

    \item[(B)] {For any morphism $f:B \rightarrow A$, and any
proposition $[A,d]$, the corresponding $B$-proposition
$[B,f^\dagger(d)]$ is intuitively logically weaker than (a
consequence of) $[A,d]$. Again, two remarks about this
assumption.

\begin{enumerate}
    \item To accommodate the identity morphism, and the
requirement of (A) that $({\rm id}_{A})^\dagger = {\rm
id}_{{\cal P}(A)}$, we note that `weaker' here means `strictly
weaker {\em or\/} the same as', just as `$\leq$' means `is
less than or equal to'. Similarly for `consequence'.

    \item Again: it is enough at this stage to use `logically
weaker' in an intuitive sense, so as to motivate the
requirements in (C) below. Subsection 2 will make it precise,
in terms of each object's family of propositions being a poset
and there being a comparison functor between them.
\end{enumerate}
    }

    \item[(C)] {We propose to assign to each proposition $[A,d]$
a truth-value $\nu(A,d)$. There is to be one truth-value,
called `total truth' (as against the other `partially true'
values), that is subject to the following intuitive
requirements:
\begin{enumerate}
    \item If $[A,d]$ is totally true, so are all its weakenings
(consequences) $[B,f^\dagger(d)]$; (note that since assumption
(A) required $({\rm id}_{A})^\dagger = {\rm id}_{{\cal
P}(A)}$, $[A,d]$ is one of its own weakenings, and so the
converse statement is automatic).

    \item If $[A,d]$ is partially true ({\em i.e.}, has one of
the other truth-values), it is in some intuitive sense `more
true', or `nearer being totally true', the more of its
weakenings $[B,f^\dagger(d)]$ are totally true.

    \item The truth-value $\nu(A,d)$, is to be determined by
which of the weakenings $[B,f^\dagger(d)]$ of $[A,d]$ is
totally true: determined in some way that obeys (a) and (b)
above.
\end{enumerate}
    }

\end{enumerate}

Three remarks about assumption (C). First: part (c) is perhaps
less intuitive than parts (a) and (b); but it can be motivated
by the philosophical idea that the semantic value or `content'
of a sentence is determined by the set of those of its
consequences that are true (in the usual classical two-valued
sense)---we discuss this in \cite{BI98b}. Second: part (c) can
also be defended as likely to mollify sceptics about partial
truth. For it makes the notion of partial truth tightly
controlled by the more acceptable notion of total truth: once
the maps $f^\dagger$ and the set of totally true propositions
is given, the partial truth-values of all propositions are
fixed---and fixed `individually' in that the partial
truth-value of $[A,d]$ depends only on which of {\em its\/}
weakenings are totally true. In any case, we now assume (c).
Third: one might propose as intuitive a variant of (b), namely
(b${}^\prime$): if $[A,d]$ is partially true, it is more true
({\em i.e.}, nearer total truth), the more of its weakenings
$[B,f^\dagger(d)]$ are near to total truth. But we will make
no use of this.

Given these assumptions, the intuitive argument proceeds in
two steps. First, these assumptions, especially part (c) of
(C), prompt a very natural suggestion for what $\nu(A,d)$
should be. Namely:
\begin{quote}
 (M): $\nu(A,d)$ is to be the set of those morphisms $f:B
\rightarrow A$ with the property that the corresponding
proposition, $[B,f^\dagger(d)]$ is totally true. In symbols:
\begin{equation}
    \nu(A,d) = \{f:B \rightarrow A \mid [B,f^\dagger(d)]
        \mbox{ is totally true }\}
\end{equation}
\end{quote}
(We write `(M)' for `morphisms'.) This suggestion makes
$\nu(A,d)$ determined by which weakenings of $[A,d]$ are
totally true, as required by (C) part (c): indeed, determined
very simply.

Second, one naturally asks: what is it for a proposition,
whether $[B,f^\dagger(d)]$ or $[A,d]$, to be totally true?
That is not yet settled. But again there is a very natural
suggestion, obeying parts (a) and (b) of (C). Namely:
\begin{quote}
(T): For any proposition $[A,d]$, total truth is just
$\nu(A,d)$ being the set of {\em all\/} morphisms, $f:B
\rightarrow A$, to $A$, {\em i.e.}, the principal sieve on
$A$. In symbols:
\begin{equation}
[A,d]\mbox{ is totally true if }\nu(A,d) = \downarrow\!\!A
\end{equation}
\end{quote}
(We write `(T)' for `total truth'.) This suggestion is
natural, because when taken together with (M):
\begin{enumerate}
    \item[a)] it follows that part (a) of (C) holds;

    \item[b)] it follows that part (b) of (C) holds, in a very
natural sense of the phrase `$[A,d]$ is more true', namely
that $\nu(A,d)$ is a larger subset of $\downarrow\!\!A$.
\end{enumerate}
Finally, to complete the intuitive argument: it follows
immediately from (M) and (T) taken together that $\nu(A,d)$ is
a sieve. For recall that, for any object $A$ in a category
$\cal C$, a set $S$ of morphisms to $A$ is a sieve if and only
if the pull-back along any morphism in $S$ is the principal
sieve.

\subsection{Assessing the Argument}
We will make two comments on the argument in the last
Subsection, and then describe how to make it precise using the
ideas in Section \ref{SubSec:ValandCGPre}.

First, we emphasise that the intuitive argument is not a
genuine {\em deduction\/} of valuations being sieve-valued. It
only claims that (M) and (T) (and therefore, sieve-valued
valuations) are natural, given (A) to (C). One could perhaps
get a genuine deduction of $\nu(A,d)$ being a sieve, but only
at the price of some strong premises. Indeed, the obvious
stronger premises that one might consider do not quite imply
(M) and (T); they just make them even more natural than they
were in Subsection 1. Thus suppose we added as premises, both:
\begin{quote}
(D) The truth-value $\nu(A,d)$ is some set of morphisms to
$A$;

(T) Total truth is to be just $\nu(A,d)$ being the principal
sieve on $A$. In symbols: $[A,d]\mbox{ is totally true if and
only if }\nu(A,d) = \downarrow\!\!A $.
\end{quote}
Even these two do not {\em imply\/} (M); though they make it
extremely natural to accept (M), and therefore to accept (as
in the argument in Subsection 1) $\nu(A,d)$ being a sieve.

More generally, we agree that essentially the same argument
can be given different versions; and we make no claim to the
version in Subsection 1 being the unique best balance between
premises being plausible and the inference being rigorously
deductively valid. (We admit that in philosophical argument,
we tend to weigh the former more highly, as shown by our
choice of version in Subsection 1.)

Second, we note that the fact that $\nu(A,d)$ is a sieve
implies the special case of our sieve-version of {\em FUNC\/},
viz. the case when $f:B\rightarrow A \in \nu(A,d)$. For by the
construction above:
\begin{equation}
    \nu(B,f^\dagger(d)) = \downarrow\!\!B;
\end{equation}
while by the definition of a sieve, and that fact that
$f:B\rightarrow A \in \nu(A,d)$,
\begin{equation}
    f^*(\nu(A,d)) = \downarrow\!\!B.
\end{equation}
But nothing in the argument implies our sieve-version of {\em
FUNC\/} in full generality, {\em i.e.}, the principle that
even when $f:B\rightarrow A \not \in \nu(A,d)$
\begin{equation}
    \nu(B,f^\dagger(d)) = f^*(\nu(A,d)).
\end{equation}
On the other hand, as we saw in Section \ref{SubSec:RoleFUNC},
{\em FUNC\/} can be motivated by the requirement that a
valuation determines a subobject of $\bf G$.

Finally, we round off this Section by showing how to make the
intuitive argument precise by using the notions of Section
\ref{SubSec:ValandCGPre}.  We only need to make assumptions
(A) to (C) precise: the argument then proceeds as in Section
\ref{SubSec:IntuitiveArgument}.  So, first: we can make (A)
precise by requiring that (i) each of the sets ${\cal P}(A)$
be a poset with a $0$ and a $1$; and (ii) the map $A \mapsto
{\cal P}(A)$ define a presheaf---as explained above, this
requirement is natural in view of the likely existence of
factorisations of morphisms $k:C\rightarrow A$. From now on,
we call this presheaf $\bf G$, as in Section
\ref{SubSec:ValandCGPre} (and earlier). So ${\cal P}(A)={\bf
G}(A)$ and $f^\dagger = {\bf G}(f)$. Note that the fact that
$\bf G$ is a presheaf now implies our requirement that $({\rm
id}_{A})^\dagger = {\rm id}_{{\cal P}(A)}$, {\em i.e.}, ${\bf
G}({\rm id}_{A})={\rm id}_{{\bf G}(A)}$.

Second, to make (B) precise: whenever $f:B \rightarrow A$, we
need to be able to `push-forward' a proposition $[B,{\bf
G}(f)(d)]$---or, in our other notation, ${\bf G}(f)(d) \in
{\bf G}(B)$---to ${\bf G}(A)$ for comparison of `logical
strength' ({\em i.e.}, comparison according to the partial
order $<$ in ${\bf G}(A)$) with the proposition $[A,d]$, {\em
i.e.}, with $d \in {\bf G}(A)$. And the pushed proposition is
required to be weaker ({\em i.e.}, higher in the partial
order) than $[A,d]$. So we require:
\begin{itemize}
\item There is a comparison functor in the (weak) sense of
Section \ref{SubSec:ValandCGPre}.1, {\em i.e.}, a covariant
functor ${\bf C}$ from $\cal C$ to ${\rm Set}$ with the same
action on objects $A$ in $\cal C$ as has $\bf G$: ${\bf
C}(A):={\bf G}(A)$.

\item The functors $\bf C$ and $\bf G$ together obey generalized
coarse-graining, Eq. (\ref{Def:CoarseGrain}), {\em i.e.}:
\begin{equation}
    d \leq {\bf C}(f)[{\bf G}(f)(d)].
\end{equation}
\end{itemize}
The generalized retraction and monotonicity clauses in the
definition in Section \ref{SubSec:ValandCGPre} of a
generalized coarse-graining presheaf are not needed {\em a
priori\/}, although we note that the monotonicity condition is
particularly natural.

Finally, (C) can be rendered precise by requiring that the set
of truth-values be a poset with a $1$ (representing `totally
true'); and that assignments of truth values, $\nu$, should
obey the following conditions:
\begin{enumerate}
\item[(a)] If $\nu(A,d)=1$, then for any $f:B\rightarrow A$,
$\nu(B,{\bf G}(f)(d))=1$.

\item[(b)] Suppose $[A,d]$ and $[A,e]$ are such that whenever
$f:B\rightarrow A$, if $\nu(B,{\bf G}(f)(d))=1$, then also
$\nu(B,{\bf G}(f)(e))=1$. Then:
\begin{equation}
    \nu(A,d) < \nu(A,e).
\end{equation}

\item[(c)] $\nu(A,d)$ is determined by the set $\{[B,{\bf
G}(f)(d)]\mid f:B\rightarrow A\mbox{, and }\nu(B,{\bf
G}(f)(d))=1 \}$.
\end{enumerate}
Given these assumptions, the argument for sieve-valued
valuations, {\em i.e.}, for the set of truth-values at each
$A$ in $\cal C$ being ${\bf\Omega}(A)$, can proceed just as in
Section \ref{SubSec:IntuitiveArgument}, though with the new
notation, $\bf G$, $\bf C$, etc.

\section{Conclusion}
To conclude, let us summarize some of our main proposals (both
in {\I} and this paper), referring mainly to the physical
cases (quantum and classical).

\noindent (1) We consider the set of physical quantities as a
mathematical category, with morphisms given by
coarse-graining, {\em i.e.}, taking functions of quantities.
So in quantum theory, we consider the category $\cal O$ of
bounded self-adjoint operators on a Hilbert space $\cal H$,
with a morphism from one such operator, $\hat B$, to another,
$\hat A$, whenever $\hat B$ is a function of $\hat A$.
Correspondingly, for classical physics, we consider the
category $\cal M$ of real-valued measurable functions on a
classical state space $\cal S$, with a morphism from one such
function, $\bar B$, to another, $\bar A$, whenever $\bar B$ is
a function of $\bar A$.

\noindent (2) We assign to each proposition, `$A \in \Delta$',
(that says the value of the quantity $A$ lies in the Borel set
$\Delta$) as its value: a sieve on $A$---a sieve on $A$ being
a set of morphisms to $A$, $f: B \rightarrow A$ that is closed
under further coarse-graining. (Here and in what follows, we
use `$A$' to stand indifferently for a quantum or classical
quantity, represented by $\hat A$ or $\bar A$ respectively.)

\noindent (3) The previous paper motivated this proposal, for
quantum theory, by linking the Kochen-Specker theorem to the
theory of presheaves.  For our category ${\cal O}_d$ of
discrete-spectrum operators, the theorem states that if the
dimension of $\cal H$ is greater than 2, then there are no
real-valued functions $V$ on ${\cal O}_d$ that have the {\em
FUNC\/} property,
\begin{equation}
    V(f(\hat{A})) = f (V(\hat{A})).     \label{V(f(A))=f(V(a))}
\end{equation}
On the other hand, a presheaf is an assignment, to each object
in a category, of a set, such that the sets assigned to
objects that are related by a morphism, `mesh' with each other
by having a corresponding set-morphism, {\em i.e.}, a
function, between them.

The Kochen-Specker theorem turns out to be a statement about
the presheaf on $\cal O$ that assigns to each operator, its
spectrum: the meshing of this presheaf turns out to be very
closely related to the meshing of values given by Eq.\
(\ref{V(f(A))=f(V(a))}). Namely, the Kochen-Specker theorem
says that this presheaf has no global elements; where a global
element is the analogue, for presheaves, of the ordinary idea
of an element of a set.

This situation suggests partial valuations on ${\cal O}_d$,
{\em i.e.}, real-valued functions on a subset of ${\cal O}_d$
that obey Eq.\ (\ref{V(f(A))=f(V(a))}); and this led us to our
proposed sieve-valued valuations on all of $\cal O$. These
have a corresponding {\em FUNC\/} property (expressed in terms
of pull-backs of sieves) and other natural properties (like
Null proposition, and monotonicity); and yet they are defined
on {\em all\/} quantities.

\noindent (4) In this paper, we have motivated these proposals
in three other ways. First, we showed that they apply equally
well to classical physics: in the absence of Kochen-Specker
prohibitions, we considered how to define a valuation given
only a macrostate (Section \ref{SieveValClassl}). Second, we
showed how some of our main proposals carry over directly to
the very general setting of any presheaf of propositions on
any small category: {\em e.g.\/} the equivalence of {\em
FUNC\/} to a sieve-valued valuation specifying a subobject
(Section \ref{Sec:GenPropSieveValuations}). Third, we showed
that our sieve-valued valuations are a very natural way to
satisfy some general intuitive requirements about partial
truth, as applied to a presheaf of propositions defined on any
(small) category (Section \ref{Sec:LogicPartialTruth}). Here
we emphasised the point that for our valuations, the partial
truth-value of a proposition is determined by which of its
weakenings are totally true.

\section*{Acknowledgements}

Chris Isham is most grateful to the Mrs L.D.~Rope Third
Charitable Settlement for financial assistance during the
course of this work.


\begin{thebibliography}{1}

\bibitem{IB98a}
C.J. Isham and J.~Butterfield.
\newblock A topos perspective on the {K}ochen-{S}pecker theorem:
{I.} {Q}uantum  states as generalised valuations.
\newblock {\em Int. J. Theor. Phys.}, 1998.
\newblock to appear. quant-ph/980355.

\bibitem{BI98b}
J.~Butterfield and C.J. Isham.
\newblock Philosophical implications of generalised valuations.
\newblock 1998.
\newblock {I}n preparation.

\bibitem{BI98c}
J.~Butterfield and C.J. Isham.
\newblock A topos perspective on the {K}ochen-{S}pecker theorem:
{III.} {I}nterval-valued generalised valuations.
\newblock 1998.
\newblock {I}n preparation.

\bibitem{Gol84}
R.~Goldblatt.
\newblock {\em Topoi: The Categorial Analysis of Logic}.
\newblock North-Holland, London, 1984.

\bibitem{MM92}
S.~Mac{L}ane and I.~Moerdijk.
\newblock {\em Sheaves in Geometry and Logic: {A} First
Introduction to Topos Theory}.
\newblock Springer-Verlag, London, 1992.

\bibitem{Haa80}
S.~Haack.
\newblock Is truth flat or bumpy?
\newblock In D.H. Mellor, editor, {\em Prospects for
Pragmatism}, pages 1--20. Cambridge University Press,
Cambridge, 1980.

\end{thebibliography}
\end{document}